\documentclass[acmsmall,screen,preprint,nonacm,natbib=false]{acmart}
\usepackage{fancyhdr}
\AtBeginDocument{%
    \addtolength{\footskip}{2.0\baselineskip}%
    \fancyfoot[L]{\textit{\textbf{Working Paper}}}%
}

\AtBeginDocument{%
  }

\setcopyright{acmlicensed}
\copyrightyear{2025}
\acmYear{2025}
\acmDOI{XXXXXXX.XXXXXXX}

\acmConference[CHI '25]{CHI conference on Human Factors in Computing Systems}{April 26 -- May 01,
  2025}{Yokohama, Japan}
\acmISBN{978-1-4503-XXXX-X/18/06}

\usepackage[
  datamodel=acmdatamodel,
  sortcites
  ]{biblatex}
\usepackage{graphicx}
\usepackage{tabularx} 
\usepackage{subcaption}
\usepackage{amsmath}
\usepackage{amsfonts}
\usepackage{booktabs}

\begin{document}

\title{``Diversity is Having the Diversity'': Unpacking and Designing for Diversity in Applicant Selection}

\author{Neil Natarajan}
\email{neilenatarajan@gmail.com}
\orcid{}
\affiliation{%
    \institution{University of Oxford}
    \city{Oxford}
    \country{United Kingdom}
}

\author{Sruthi Viswanathan}
\email{sruthi.viswanathan@cs.ox.ac.uk}
\orcid{}
\affiliation{%
  \institution{University of Oxford}
  \city{Oxford}
  \country{UK}
}

\author{Reuben Binns}
\email{reuben.binns@cs.ox.ac.uk}
\orcid{}
\affiliation{%
    \institution{University of Oxford}
    \city{Oxford}
    \country{United Kingdom}
    \postcode{OX1 2JD}
}

\author{Nigel Shadbolt}
\email{nigel.shadbolt@cs.ox.ac.uk}
\orcid{}
\affiliation{%
    \institution{University of Oxford}
    \city{Oxford}
    \country{United Kingdom}
    \postcode{OX1 2JD}
}

\renewcommand{\shortauthors}{Natarajan et al.}

\begin{abstract}
    When selecting applicants for scholarships, universities, or jobs, practitioners often aim for a diverse cohort of qualified recipients. However, differing articulations, constructs, and notions of diversity prevents decision-makers from operationalising and progressing towards the diversity they all agree is needed . To understand this challenge of translation from values, to requirements, to decision support tools (DSTs), we conducted participatory design studies exploring professionals' varied perceptions of diversity and how to build for them. Our results suggest three definitions of diversity: bringing together \emph{different perspectives}; \emph{ensuring representativeness} of a base population; and \emph{contextualising applications}, which we use to create the \emph{Diversity Triangle}. We experience-prototyped DSTs reflecting each angle of the \emph{Diversity Triangle} to enhance decision-making around diversity. We find that notions of diversity are highly diverse; efforts to design DSTs for diversity should start by working with organisations to distil `diversity' into definitions and design requirements.
\end{abstract}



\received{12 September 2024}
\received[revised]{12 September 2024}
\received[accepted]{12 September 2024}

\begin{teaserfigure}
  \includegraphics[width=\textwidth]{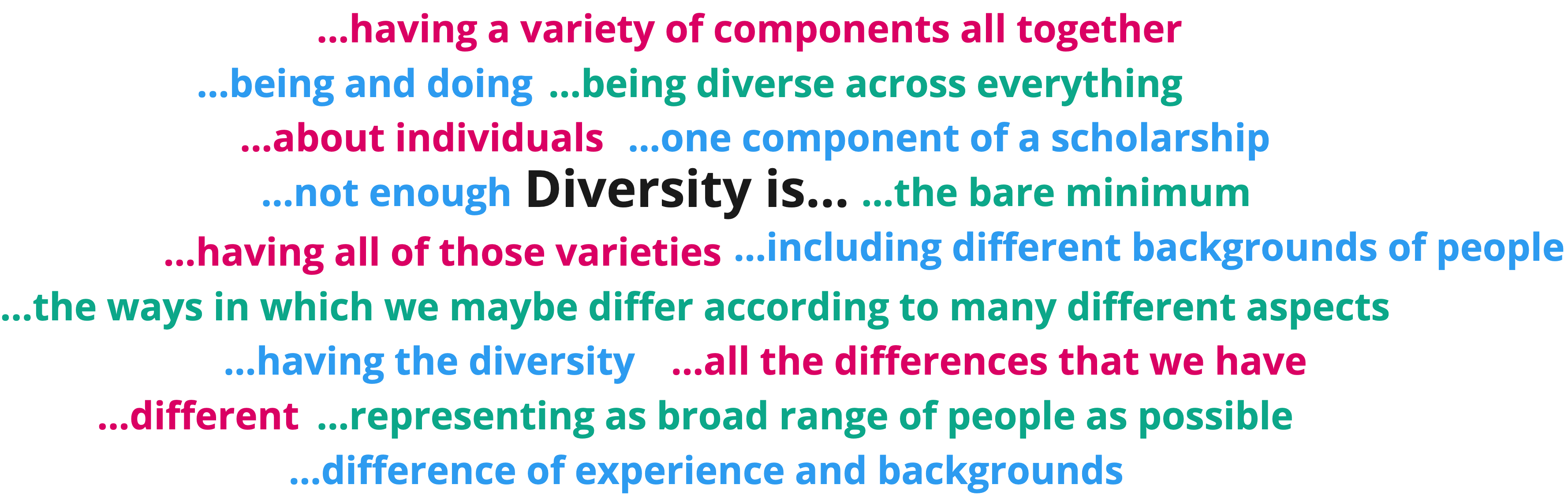}
  \caption{This figure shows participant codes defining what ``diversity is''. In this paper, we seek to answer: what do they mean and how do we design for that?}
  \Description{Participant definitions of diversity.}
  \label{fig:diversity_is_teaser}
\end{teaserfigure}

\maketitle

\section{Introduction}\label{sec:intro}
Processes for selecting people for jobs, universities, prizes, and other opportunities have often failed to reflect the diversity of their actual and potential candidate pool. Recognising this, various sectors have in recent years shifted towards recognising and promoting diversity through the establishment of a variety of related norms: DEI (Diversity, Equity, and Inclusion), EDI (Equality, Diversity and Inclusion), JEDI (Justice, Equality, Diversity, and Inclusion), DEIB (Diversity, Equity, Inclusion, and Belonging), etc. \cite{pinkett2023data,hsieh2019allocation,minkin2023diversity}. These norms are frequently operationalised through changes to application, evaluation, and decision-making procedures designed to result in greater representation of different demographic groups in final selection decisions \cite{pinkett2023data}. Such efforts are in part a response to widespread societal concerns about racial, gender, and other injustices, but have also frequently been justified in economic terms by evidence suggesting that diverse teams perform better than homogeneous ones on a variety of tasks \cite{deming2017growing,page2019diversity,noray2023systemic}. Meanwhile, the concept of diversity itself has become swept up in `culture war' discourse, criticised by right wing commentators as part of a sinister `woke' agenda, and by progressives as mere window dressing which fails to meaningfully address deeper societal injustices. 

However, for practitioners involved in selection processes on the ground, the question of how to meaningfully measure and promote diversity in their decisions is a real and challenging one. The proliferation of software to help with hiring and talent selection – in the form of application management platforms, decision-support tools, and more recently, widespread use of AI – provides additional complications, as well as potential opportunities \cite{Lashkari_Cheng_2023}. For instance, AI-driven tools may discriminate against candidates who don't resemble their training data \cite{chen2018investigating,li2020hiring,lambrecht2019algorithmic}; but they could potentially assist in mitigating human biases in recruitment and selection processes, resulting in greater diversity \cite{yarger2020algorithmic,avery2024does,will2023people, suhr2021does}. This context thus challenges Human-Computer Interaction (HCI) and Human-Centred Artificial Intelligence (HCAI) research to improve diversity in selection processes without amplifying or codifying pre-existing inequities.

But to improve diversity in selection processes, we must first understand it. We undertake two studies to do just that. In Study 1, we conducted 15 one-to-one interviews with scholarship and talent investment selection practitioners (selectors). All participants were involved in scholarship selection for an international cohort of students in a global academic programme. We aimed to understand how these practitioners define and operationalise diversity and how technology can aid this process. Each interview lasted 45 minutes. We analysed this data following \textcite{braun_using_2006}'s inductive thematic analysis methodology. Our analysis highlighted the ad-hoc nature of current diversity considerations and the need for more structured, data-supported approaches, and surfaced three distinct definitions of diversity: one involves placing people with `different perspectives' in the same room; another, ensuring `representativeness' of some target population; a third, `contextualising applications', e.g., with information such as the applicants' relative privilege or required level of support. We conclude that technological interventions designed to promote diversity in selection processes should first identify specific definition(s) of diversity they aim to promote; we construct the \emph{Diversity Triangle} (Figure \ref{fig:div_triangle}) as a guide.

For Study 2, we developed six design prototypes for diversity-supporting tools based on the Diversity Triangle from Study 1 \cite{Buchenau_Suri_2000}. The prototypes included tools for visualising cohort representativeness, measuring entropy (average number of in-group differences), and assessing individual applicant (dis)advantage scores. They were presented to participants via participatory design workshops \cite{Zimmerman_Forlizzi_2017}. Participants provided feedback on the utility and integration of these tools into their selection processes. The workshops revealed that participants use specific, idiosyncratic lenses to navigate their diversity considerations, and that while quantitative tools are essential for making informed decisions, qualitative assessments remain crucial.

This research contributes to the understanding of how diversity can be supported through data-driven tools in selection processes. Specifically, our contributions are:

\begin{enumerate}
    \item Three definitions of diversity that impact selectors' decisions uncovered through inductive thematic analysis.
    \item The Diversity Triangle, categorising diversity-related themes according to our definitions of diversity.
    \item Design recommendations grounded in participatory design for system implementers supporting the diversity needs of a given organisation.
\end{enumerate}

\noindent More broadly, this work demonstrates that providing structured, data-supported approaches to diversity, organisations can better navigate the complexities of DEI (EDI, JEDI, DEIB, etc.). While differing in some respects, we believe these implications will generalise from scholarship selection to various other talent identification contexts, including recruitment for jobs and admission for universities. By helping these organisations achieve their desired outcomes in selection processes, we ultimately contribute to and help build a more diverse society.

\section{Background}\label{sec:back}
\subsection{Diversity as a societal and organisational value}\label{ssec:value}
Despite its global reach, contemporary discourse on diversity derives (in large part) from the political context of the United States in the latter half of the 20th century \cite{nkomo2019diversity}. Civil rights activists identified gender, race, disability, and other forms of group identity as loci of discrimination and oppression, and constructed political actions around these identities \cite{morris1984origins}. This yielded civil rights laws, including equal treatment laws to protect applicants from discrimination (e.g., in employment). \textcite{nkomo2019diversity} argue that this initial, U.S.-centric perspective on anti-discrimination in the workplace, which focused on under-representation of women and racial minorities, has evolved into a more global concept of diversity encompassing a variety of identities. With an increase in social pressure for representation, organisations increasingly prioritise diversity in their selection procedures \cite{hsieh2019allocation,minkin2023diversity}.

But social and political pressures do not uniformly push for diversity. Critics on all sides challenge the value of diversity, both to organisations and society. \textcite{Ahmed_2012} argues that organisational prioritisation of diversity often limits their appetite to prioritise more meaningful changes. The attention paid to diversity may encourage organisations to merely document social injustice, rather than do something to change it \cite{Ahmed_2012,Rossi2020-ROSWNA-2}. Worse, \textcite{Warikoo_2019} argues that diversity among elite institutions reinforces social injustices. On the other side, critics such as \textcite{Goodhart} position diversity as opposed to the obligations of a good society, while others position this as less a rejection of `diversity' than a rejection of `bad diversity' \cite{lentin_Multiculturalism_2011}.

However, political struggles and social pressure for greater equality are not the only motivation for pursuing diversity. More recent discourse and research has also emphasised the instrumental benefits that greater diversity can bring to workplaces and institutions. While it is sometimes assumed that diversity could reduce the quality of a workforce or cohort by favouring more diverse but `worse' candidates, research on selection processes suggests many organisations can both improve the diversity of their organisations and select `better' candidates (as measured by their own metrics) \cite{autor2008does,noray2023systemic}. Generally, research on the performance of diverse teams yields mixed results \cite{daubner2017dovetailing,page2019diversity,noray2023systemic,muller_learning_2019}.

Other arguments in favour of diversity emphasise its benefits in the context of knowledge production. For instance, a long strand of work in philosophy of science and social epistemology has made the case for diversity as key to success in epistemic communities \cite{mill1998liberty,merton1942note,wylie2006introduction}. While diversity in this context often refers to cognitive diversity rather than demographic diversity, the latter may often precede the former, as: ``membership in different social groups (e.g., gender or race) often comes with different task-relevant information, perspectives, or experiences'' \cite{peters2021hidden}. For instance, in biomedical science, inclusion of scientists from more diverse backgrounds could lead to ``novel findings and treatment of diverse populations'' \cite{swartz2019science}.

More generally, ``the more robust a community’s mechanisms for bringing diverse perspectives to bear on epistemic questions of public import, the more effective it is in generating and ensuring responsiveness to all the information and insights held by its members'' \cite{wylie2006introduction}. A related argument is that marginalised groups may occupy epistemically advantageous standpoints when it comes to understanding unjust power structures \cite{harding2004feminist,dror2023there,steel_multiple_2018}, as they may have more informative experiences of oppression \cite{mills2015blackness}, and incentives to learn about it \cite{jaggar1983feminist}.

\subsection{Defining and Measuring Diversity}
With the wealth of disparate motivations for diversity, it is unsurprising that its definition is often unclear. \textcite{page2010diversity} offers a helpful generic definition of diversity: ``The heterogeneity of elements in a set in relation to a class that takes different values, such as species in an eco-environment, or ethnicity in a population''. While suitably broad, this definition lacks the specificity required to build supporting technologies \cite{hupont2021diverse,page2010diversity}. 

However, different means of measuring diversity appear to offer vastly differing accounts of specific definitions. For instance, the natural way to measure diversity, on \textcite{page2010diversity}'s definition, is to report percentages of those different elements, e.g., demographics in a population. But in scientific fields like biology and ecology, diversity is often measured with different methods and occasionally conflicting results \cite{xu2020diversity}. Some use measures from information theory (including entropy measured through the Shannon index \cite{shannon1948mathematical}); others adapt the Herfindahl-Hirschman Index \cite{rhoades1993herfindahl}, commonly used to measure economic market concentration \cite{budescu_how_2012,acuna2021ai,shannon1948mathematical,rhoades1993herfindahl}.

Another complication arises when we consider the relationship between diversity and similar notions from AI ethics like fairness: are these really distinct concepts, or do they collapse into each other?
\textcite{zhao2023fairness} argue that ``fairness works can be re-interpreted through the lens of diversity, and strategies enhancing diversity have proven efficacious in improving fairness''. If we assume this re-interpretation, algorithmic fairness literature implies another approach to measuring diversity in constraining model outputs to equalise the performance (i.e., positive predictions) between different demographic groups \cite{barocas2023fairness}. And, though many of these measurements are defined for non-human notions of `diversity' (e.g., diversity of items in a recommender system or diversity of flora in an ecosystem), HCI and related research communities often apply these measurements to human notions when measuring the diversity of participants for research, records for AI training data or backgrounds of authors of published papers \cite{linxen2021weird,himmelsbach2019we,zhao2024position,rojas2022dollar,acuna2021ai}. 

More practical criticisms of measurement and implementation of notions of diversity often point to this confusion. \textcite{steel_multiple_2018} argue that disparate definitions of diversity: ``Can generate unclarity about the meaning of diversity, lead to problematic inferences from empirical research, and obscure complex ethical-epistemic questions about how to define diversity in specific cases''. Similarly, \textcite{abdu2023empirical} caution that the specifics of measurement are, at least in the context of race: ``a value-laden process with significant social and political consequences''. And, if done poorly, this measurement process may exclude already marginalised groups (e.g., gender-fluid or mixed-race individuals) \cite{scheuerman2019computers}. While poor categorisation is harmful in any context, larger institutions must be doubly cautious, as, in using constructs to measure diversity, they may inadvertently reify them as natural rather than social \cite{scheuerman2021auto}. 

\subsection{AI and Decision Support Systems for Hiring and Talent Selection}
While there is good reason to express scepticism towards predictive technology in recruitment, \textcite{Vereschak_Alizadeh_Bailly_Caramiaux_2024} demonstrate that, so long as AI systems prove trustworthy and well-designed, decision-makers will engage with and rely on these systems. Apprehension with these systems range from concerns about bias to a disquiet about a distance between applicant and organisation \cite{Lashkari_Cheng_2023}. While concerns about biased AI systems can also be directed at human reviewers, the distance between applicant and organisation created by the introduction of automated decision support tools must be more carefully managed \cite{Leung_Zhang_Jibuti_Zhao_Klein_Pierce_Robert_Zhu_2020,Lashkari_Cheng_2023}. Ultimately, flaws in present pipelines point to the need for better systems, and, despite a well-earned weariness of technologies developed for this sensitive field, AI and decision-support technologies may have a role to play \cite{kleinberg2018algorithmic,Vereschak_Alizadeh_Bailly_Caramiaux_2024,barocas2023fairness,huppenkothen2020entrofy,schumann_diverse_2019}.

New research explores a framing of individual applicant aptitude and overall group diversity \cite{noray2023systemic}, and many applications demonstrate that technology might improve both. For example, \textcite{bergman2021seven} show that replacing traditional testing mechanisms with prediction algorithms allows placement of students into remedial classes that improves student performance and increases minority representation in non-remedial courses. Similarly, \textcite{autor2008does} show that screening job applicants with personality tests increases worker productivity without reducing minority representation. Qualitative studies of HR practitioners' use of AI and decision-support systems have found that diversity is among the perceived benefits of such systems; \textcite{li2021algorithmic} find that recruiters and HR practitioners who used AI-enabled hiring software for sourcing and assessment reported higher diversity in their candidate pools.  \textcite{huppenkothen2020entrofy,kleinberg2018algorithmic,schumann_diverse_2019} demonstrate algorithms directly comparing applicant aptitude to group diversity, and all find improvements on both axes over traditional selection procedures. However, despite these widespread improvements, these technologies are rarely used in practice \cite{page_diversity_2017}, as the processes assumed by these technologies rarely align with that of selection organisations. More work is needed that begins with practitioners' processes and diversity needs, then constructs data-based support systems from there.

\section{Experimental Design}\label{sec:methods}
\subsection{Our Studies}
\begin{figure}[htbp]
    \centering
    \includegraphics[width=.9\textwidth]{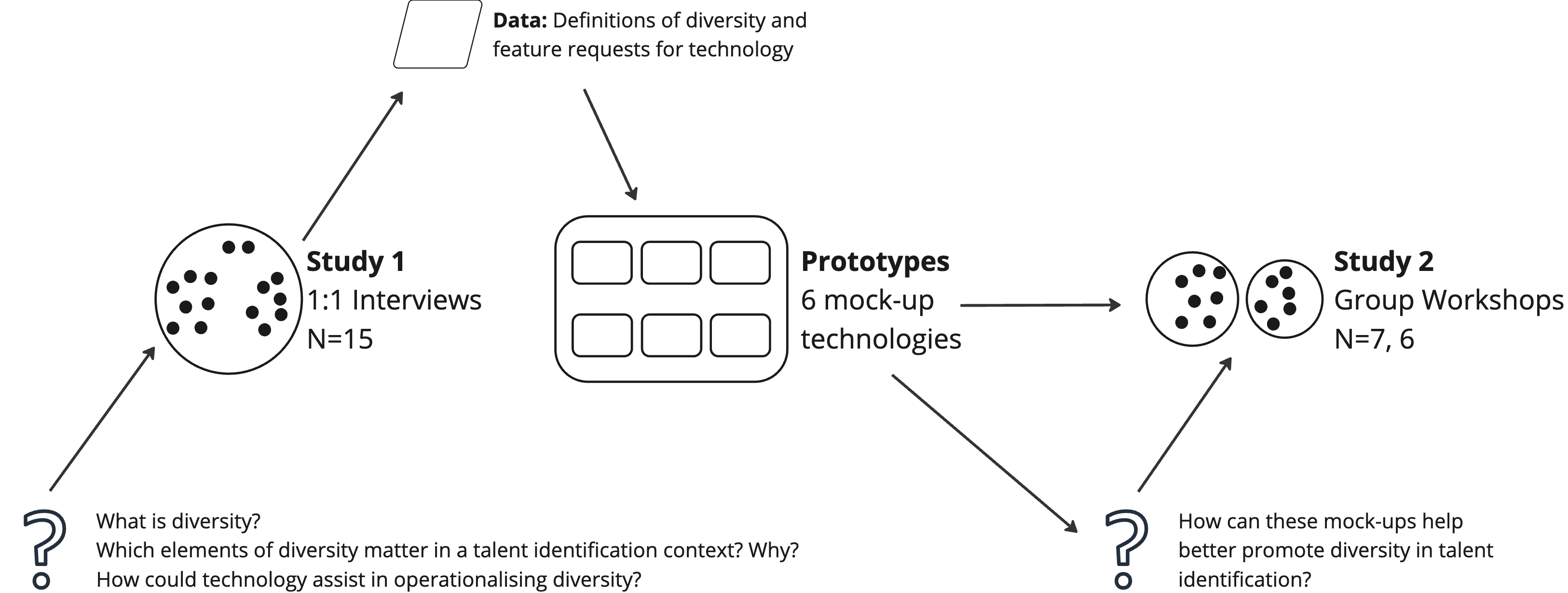}
    \caption{This research begins with 15 interviews seeking to understand what selection practitioners mean when they talk about diversity and how to support that, followed with two scenario-speed-dating activities where these practitioners test a number of prototypes built based on the interviews.}
    \label{fig:flowchart}
    \Description{A Flowchart depicting the relationship between studies.}
\end{figure}

Our research is broken into two components: 15 one-to-one interviews with scholarship and talent investment selection practitioners and two participatory design workshops with a subset of the 15 practitioners. The relationship between these components is shown in Figure \ref{fig:flowchart} and details on Studies 1 and 2 can be found in Sections \ref{ssec:methods1} and \ref{ssec:methods2}, respectively.

In the first session, we seek to ascertain how these practitioners understand diversity, how they operationalise it in processes for selecting talented applicants, and how they envision using technology to assist them in that process. In the second session, we show these practitioners six prototypes built in response to the interviews, then we aim to collaboratively design tools that can help them better consider diversity in selection.

\subsection{Positionality}
Following \textcite{Venn-Wycherley_Kharrufa_Lechelt_Nicholson_Howland_Almjally_Trory_Sarangapani_2024}, we state researcher positionality here. All authors endorse diversity as a societal and organisational value as discussed in Section \ref{ssec:value} (while sympathetic to the critiques of \textcite{Ahmed_2012,Warikoo_2019}); we thus contend that improving organisational capability to consider diversity is generally a positive development. The research team is comprised of [researcher demographics redacted for review].

\subsection{Participants}
We engaged selection practitioners (N=15) from two scholarship and talent investment programs. These participants' numbers can be seen in Table \ref{tab:participants}. Where participants participate in the second activity, their groupings for the activities can be seen here as well.

Before either study, we obtained informed consent from all participants to be included in both studies. All participants were given the option to recuse themselves from Study 1 at any point until publication, but, as Study 2 is a group workshop (and as we do not do participant-level attribution), participants were asked to recuse themselves before this second study. Two participants recused themselves before Study 2, but gave leave to be included in Study 1. Participants also gave consent to be recorded, and to have these recordings stored on a secure server. All recording, transcribing, and data analysis was conducted on secure servers. Ethics review was performed by [REDACTED FOR REVIEW].

\begin{center}
\begin{table}[htbp]
    \centering
    \caption{Our fifteen participants are interviewed then split between two groups, with two participants in neither group activity.}
    \label{tab:participants}
    \begin{tabular}{l l l}
        \toprule
        Activity Group 1 (G1) & Activity Group 2 (G2) & No Activity Group (N/A)\\
        \midrule
        P1 & P2 & P8 \\
        P4 & P3 & P12 \\
        P5 & P6 & \\
        P9 & P7 & \\
        P10 & P11 & \\
        P13 & P15 & \\
        P14 & & \\
        \bottomrule
    \end{tabular}
\end{table}
\end{center}

\section{Part 1: Interviews and Thematic Analysis}\label{sec:study1}
\subsection{Methodology}\label{ssec:methods1}
Our interviews aim to answer three questions in each organisational context:

\begin{enumerate}
    \item What is diversity?
    \item Which elements of diversity matter in a selection context? Why?
    \item How could technology assist in operationalising diversity?
\end{enumerate}

In answering our first set of research questions, we follow \textcite{braun_using_2006}'s methodology for reflexive thematic analysis. We first conduct 45-minute semi-structured interviews with 15 selection practitioners. In each interview, we first ask general questions about their selection methodology; we next ask specifically about diversity and its role in selection; we move on to \textcite{Knapp_Zeratzky_Kowitz_2016}'s `crazy 8s' exercise, where participants give eight feature requests in eight minutes; we conclude with a `magic app' exercise inspired by \textcite{blythe2014research}'s design fiction, where participants more thoroughly detail their ideal app. A question-by-question protocol for these interviews is supplied in Appendix \ref{app:protocol1}.

The lead author interviewed participants, then transcribed and anonymised the interviews. The lead author and another author (who wasn't present during the interviews) then independently `open-coded' each anonymised transcript to mitigate bias, looking for anything relevant to our research questions. The researchers met six times to discuss their open codes, then shared these codes with the remainder of their research team across four meetings; the researchers grouped codes into 6 themes and 18 subthemes after consensus was reached. These themes are detailed in Table \ref{tab:themes} and described in Section \ref{ssec:themes}.

\begin{table}[htbp]
    \centering
    \caption{Our three central subthemes all speak to the question ``why diversity?''. Other themes and subthemes reflect types of diversity, concepts intertwined with diversity, and other considerations scholarship programs must weigh against diversity desires.}
    \label{tab:themes}
    \begin{tabular}{|p{0.45\textwidth}|p{0.45\textwidth}|}
        \hline
        \multicolumn{2}{|c|}{\textbf{Themes and Subthemes}} \\
        \hline
        \textbf{Why Diversity?} & \textbf{Types of Diversity} \\
        \underline{Different perspectives} & \underline{Socioeconomic} \\
        \emph{...in the same room} & \emph{parental income} \\
        \underline{Representativeness} & \emph{parental education} \\
        \emph{...of a general population} & \emph{generational wealth} \\
        \emph{...of the eligible population} & \underline{Sex, gender, and sexuality} \\
        \emph{...of the applicant population} & \emph{sex} \\
        \emph{...of a target population} & \emph{gender identity} \\
        \underline{Contextualising applications} & \emph{sexual orientation} \\
         & \underline{Geography} \\
         & \emph{nationality} \\
         & \emph{the `Global South'} \\
         & \emph{region} \\
         & \underline{Race} \\
         & \emph{international categorisations of race} \\
         & \underline{Types of thinking} \\
         & \emph{subject area interest} \\
         & \emph{personality type} \\
         & \emph{core beliefs} \\
         & \emph{problem solving approaches} \\
         & \emph{political views} \\
        \hline
        \textbf{Operational Risks and Considerations} & \textbf{Fairness and Bias} \\
        \underline{Outreach} & \underline{Fairness} \\
        \underline{Support} & \emph{...to the applicants} \\
        \emph{...during the application process} & \emph{...to the world} \\
        \emph{...after selection} & \underline{Bias} \\
        \underline{Selectors} & \emph{measurement bias} \\
        \underline{Applicant fraud} & \emph{decision-makers' bias (prejudicial)} \\
         & \emph{decision-makers' unique perspective (probative)} \\
        \hline
        \textbf{Scholarship Goals} & \textbf{Merit} \\
        \underline{Impact} & \underline{Performance relative to disadvantage} \\
        \emph{...on all applicants} & \underline{Measurement} \\
        \emph{...on the selected scholars' performance} & \underline{Performance} \\
        \emph{...on the selected scholars' opportunities} & \\
        \emph{...by the selected scholars on the world} & \\
        \hline
    \end{tabular}
\end{table}

\subsection{Themes}\label{ssec:themes}
\subsubsection{Why Diversity}
A central theme of our investigation revolved around the question of why diversity matters. To this end, we asked questions such as: ``What is diversity?'' and ``Why does diversity matter?''. Answers to: ``What is diversity?'' are visualised in Figure \ref{fig:diversity_is_teaser}; as hypothesised, these answers are vague and uninformative. Interestingly, though, answers to: ``Why does diversity matter?'' informed more specific definitions. We were able to cluster these more specific definitions into three central subthemes: `different perspectives', `representativeness', and `contextualising applications'. These are listed as subthemes of `Why Diversity' in Table \ref{tab:themes}.

\paragraph{Different Perspectives}
8 out of 15 participants mentioned that diversity was important because it brought different perspectives into the same room. This was seen as important for a few related reasons, e.g., the ability to see problems from different angles and the ability to make better decisions. Several participants referred to the: ``Benefits of diverse perspectives'' (P1). One said, when discussing their personal experience working with winners in a talent investment program, that there is: ``Magic happening with lots of...diverse perspectives in the room'' (P14). Peoples' experiences were particularly relevant here. As one participant writes: ``You want to have diverse perspectives from people who look different with different experiences'' (P2).

\paragraph{Representativeness}
9 out of 15 participants spoke of `representativeness'. This, we observed, was often spoken of in relationship to a larger population. Most frequently, participants spoke of the importance of having a cohort that was representative of the `eligible population'. I.e., one participant said: ``[You want] a community which is representative of where you are selecting young people from'' (P6). Others spoke of this in broader, more general terms: ``[You want] as broad a range of people as possible'' (P15). Participants identified the importance of building a cohort that variously: ``Reflects the population of the countries'' (P15) and is ``More representative of the national population than the STEM field already is'' (P10). Others talk about representation of a particular target population, i.e.: ``Representation...because that gives you insight for the people that you're trying to serve....you have to be....reflective of your market'' (P9). Finally, selectors discussed the importance of representing an applicant population: ``[We want] a cohort that is representative of the pool'' (P10).

\paragraph{Contextualising Applications}
`Contextualising applications' was often spoken of by 6 out of 15 participants, most often in individual terms, speaking of identifying particular applicants in need of support, then offering them a `boost' in the form of said support. I.e.: ``Identify those talents and specifically boost up people who are in need of support'' (P5). One participant identified `boosting' as a key metric for the programme: ``We need to know that...we have some level of...impact here, and...if all we're doing is supporting someone who is already on an amazing trajectory and then maybe that means we're not altering their trajectory at all. That's a question of efficiency of our dollars'' (P8).

In some cases, the need for support or boosting was identified with underrepresented or disadvantaged demographic groups. One participant said: ``The focus on gender has been to give the sex that has had the least opportunity the opportunity in this program'' (P9).

\subsubsection{Types of Diversity}
Another central focus of our investigation was on different types of diversity. We asked participants to break down their understanding of diversity into different elements, and to discuss why these elements were important. We found that participants identified a wide range of different types of diversity, which we clustered into subthemes. These subthemes are listed as subthemes of `Types of Diversity' in Table \ref{tab:themes}.

Notably, in addition to the standard demographic categories commonly considered `demographic diversity', participants identified a wide range of other types of diversity commonly termed `cognitive diversity' \cite{page2010diversity}. These included `subject areas of interest', `personality type', `core beliefs', `problem solving approaches', and `political views'.

\paragraph{Socioeconomic}
All 15 participants identified socioeconomic diversity as particularly important in the context of a talent investment program. One participant said: ``Socioeconomic [diversity] is probably the most important'' (P1). Another said: ``Socioeconomic background is number one from my perspective'' (P5). This was identified as particularly important for several reasons. Participants stated: ``Because right now the SAT, for example, is more highly correlated to socioeconomic status than it is to anything else'' (P5), and: ``It's a scholarship scheme, so I think it should be for kids who cannot afford normally the fees at the university'' (P7).

Outside of the standard categorisations by income and wealth, participants also identified: ``Familial education level'' (P5) or ``Socioeconomic backgrounds'' (P10) as a particularly important metric for understanding socioeconomic diversity in a scholarship context. This suggests that historical socioeconomic status is considered alongside present socioeconomic status. One participant noted that socioeconomic status varied in both meaning and measurement from country to country: ``For example, in Columbia, there's a whole society to organise on a 1 to 7 scale for socioeconomic status'' (P5). (Columbia's policy of socioeconomic stratification divides households into six strata; unlike traditional measurements such as income or quality of life, these strata only consider household location and accommodation and thus capture a different facet of socioeconomic status \cite{CHICAOLMO2020102560}.)

\paragraph{Sex, gender, and sexuality}
While all 15 participants noted some manner of sex, gender, and sexuality diversity as important, participants disagreed on the relative emphasis that should be place on each. One participant noted: ``[Sex] is important. I think it will get diluted if we focus on identity gender because....the purpose of diversity on the gender aspect was to make sure that [men and women] were getting equal opportunities'' (P9). Another noted that, while sexual identification diversity was important in other contexts, they ``Wouldn't select for that'' (P1) in this context. Others listed `sexual orientation' and `gender' as important metrics for understanding diversity in a scholarship context. However, save for the participant who noted the distinction between sex and identity gender, participants expressed reluctance to discuss the relationships between these difficult concepts.

\paragraph{Geography}
14 of the 15 participants mentioned the importance of geography, citing a need for: ``[A] wide array of different geographical...representations'' (P8). Others spoke of a ``Regional distribution'' (P1), which we have included here.

In particular, emphasis was placed on geographic markers of socioeconomic status such as  the `Global South', `indigenous communities', and `low income countries'. One participant noted: ``Immigration status is tied so closely to socioeconomic status'' (P2), while another noted that ``[Geography] is connected to socioeconomics because we know there are some poorer countries and rich countries'' (P7). Others still asked questions like: ``Do they have a passport?....Are they in a refugee camp?'' (P5).

Furthermore, participants saw it as important that their programmes had: ``Global reach'' (P7). They expressed a desire for: ``[A] diversity of people coming from variety of places'' (P7).

\paragraph{Race}
While 11 out of 15 participants identified race as an important dimension of diversity, none suggested they would explicitly select for racial diversity. Several participants instead noted the difficulty of measuring race in a global context: ``Racial categories obviously vary by country'' (P10). One participant noted: ``[In places] like Brazil or England, there are different categories of race than there are in the US...[In Brazil] there's a board of people who decide what people's race are'' (P5). In a global context, however, many participants pointed to relationships between geography and race, and hence suggested diversifying across geography in place of race: ``If it's an international programme then you can use geography as a proxy'' (P5).

\paragraph{Types of Thinking}
4 out of 15 participants discussed a diversity in the types of thinking exhibited by applicants. One participant noted that: ``You want as much representation from different different types of thinking as you can, because I want perspectives to be listened to equally'' (P2). This manifested in many ways.

Participants tended to express the belief that personality-type-diversity could improve group cohesion: ``With that understanding [of] personality types...be able to tell which...people would get on well with each other'' (P14). One participant suggested a ``Personality test'' (P12), and another specifically mentioned a desire to diversify across ``Openness'' (P2).

However, while personality type was seen as important, core beliefs were seen as even more so. One participant noted an interest in diversity of ``Interests politically'' (P12), and expressed a desire for diversity of ``People's core beliefs...separate from religion'' (P12). Another also noted: ``I would try to have a good representation of...religious groups'' (P3).

\subsubsection{Operational Risks and Considerations}
While our study did not focus on the operational aspects of selection, several selectors' understanding of diversity was closely tied to the operational realities of selecting for and running a scholarship. In answering our questions, several participants identified operational risks or considerations that impacted their understanding of diversity. These are listed as subthemes of `Operational Risks and Considerations' in Table \ref{tab:themes}.

\paragraph{Outreach}
While our study was focused primarily on selecting a diverse cohort from a fixed applicant pool, 6 out of 15 participants answered questions from the perspective of outreach with the goal of growing a more diverse pool of applicants to select from. In particular, participants suggested that ``Using technology for....targeted outreach'' (P4) could help improve overall cohort diversity before selection even begins. One said: ``Giving you very clear signposting on where you may want to focus, you know further recruitment or outreach or whatever it might be to make sure that your programme is diverse at the end of the day'' (P6). Another added: ``You can target your outreach dollars to communities where you know that underrepresented talent exists'' (P10).

\paragraph{Support}
Similarly, 8 out of 15 participants suggested that technology could enable the support of applicants from underrepresented groups, which would also improve diversity. One participant suggested that technology could be used to provide: ``[Support] to keep people that you're attracting from underrepresented backgrounds and help them get across the the finish line'' (P10). Participants focused on the: ``Support needed to actually get [applicants] through your programme'' (P10), i.e., supporting applicants after acceptance. Another suggested that technology could be used to provide support to applicants ``After selection'' (P15).

\paragraph{Selectors}
4 out of 15 participants noted that diversity did not apply merely to applicants. Instead, for programs where a group of selectors assists in the selection process, ``Tracking the diversity of the selectors'' (P15) and ``[Monitoring] how they're scoring and reviewing applicants [for] prejudice or biases'' (P15).

\paragraph{Applicant fraud}
Finally, only 2 out of 15 participants expressed concern with selecting based on particular diversity characteristics, especially self-reported metrics of diversity characteristics, was the potential for applicant fraud, e.g. falsely reporting demographic or other attributes with the aim of increasing their chances of acceptance. One participant requested: ``A fraud detector'' (P5), while another expressed a desire to ensure that the process ``Isn't super gameable'' (P10).

\subsubsection{Fairness and Bias}
Though not a type of diversity as we have understood it here, many participants referenced similarities between diversity metrics and metrics of fairness or bias (as in \textcite{zhao2023fairness}). Furthermore, several suggested that improving fairness while reducing bias would likely yield a more diverse cohort. These themes are reflected under `Fairness and Bias' in Table \ref{tab:themes}.

\paragraph{Fairness}
6 out of 15 participants discussed that it was important that applicants: ``Get fair chance on their on their academic merit'' (P7). This translated to an emphasis on ``Fairness in the assessment'' (P7).

However, participants also noted that ``The way the world works is unfair'' (P14), and found it important that the programme is: ``Making sure that the world is fairer by bringing more diversity to this world'' (P7). In this way, participants found: ``[The] representative thing....goes back to fairness'' (P15). One participant noted that: ``Affirmative action....can come across as unfair to some people, but....it's trying to balance things out when things have been so unequal for so long'' (P3). This supports \textcite{zhao2023fairness}'s positioning of fairness as related to, rather than in opposition to, diversity. 

\paragraph{Bias}
11 out of 15 participants discussed bias, often as both a human- and machine-decision-making problem. Many participants appealed to technology's ability to be comparatively impartial as an important mitigator of bias, i.e., one participant repeatedly requested a: ``Non-biased program'' (P3); another stated a preference for: ``Data analysis to make decisions on who we should be supporting as opposed to having humans try to make those decisions with all their biases'' (P5). However, those same participants noted that common machine decision-making paradigms amplify bias, and that it was important to be aware of this: ``AI has a lot of bias in it'' (P3).

Others noted the possibility for technology to elucidate biases in both humans and machines. One participant requested: ``Some kind of tool that can detect bias in a selection'' (P7).

\subsubsection{Programme Goals}
Selection practitioners from both programmes identified goals for their programmes. These goals were discussed by both groups as an intended form of `impact' and both groups closely related achieving their goals to their expressions of why diversity mattered. While impact goals vary based on the type of impact and the affected party, we discuss these as `impact' in Table \ref{tab:themes}.

\paragraph{Impact}
5 out of 15 participants found key goals of their program to include: ``Orient[ing] [scholars] towards social impact or using their talent for good'' (P8). They tended to encourage: ``[Scholars'] working towards something impactful over the course of their career'' (P8).

\subsubsection{Merit}
Several participants reflected on the relationship between merit and diversity. While some participants saw these as competing goals, others saw them as complementary. In particular, complementary views often viewed merit as a form of performance relative to specific advantages, or noted that many of our measurement tools are biased across our chosen diversity dimensions. These themes are reflected as subthemes of `Merit' in Table \ref{tab:themes}.

\paragraph{Performance relative to disadvantage}
2 out of 15 participants identified merit as something difficult to disentangle from performance. One participant noted that applicants may appear less qualified because they: ``Didn't have the chance; didn't have the opportunities'' (P2), while others with the opportunities will appear more qualified. Another participant began by asking: ``How good is their three A stars based on where they've come from?'' (P12), then proceeded to reflect that ``Your performance relative to your opportunity or maybe expected performance'' (P12) is a key indicator of merit.

\paragraph{Measurement}
Closely related, 3 out of 15 participants questioned our ability to measure merit independent of opportunity: ``[Whether they perform well] because they have the opportunity or because they are brilliant – I think that these two are really difficult to untangle'' (P7). Another noted that: ``Contextual factors mess up our otherwise seemingly objective measures of merit.... national context and family income is messing up your ability to measure the thing you actually care about'' (P10). They continued to note that they: ``Need to pay attention to [diversity] because it's messing up your measures of what you actually care about'' (P10).

\paragraph{Performance}
Finally, 2 out of 15 participants noted occasions where performance and diversity were ostensibly competing goals. However, even here, participants recognised that observed performance and actual merit may differ. One participant noted that: ``[The] overriding aim is for [the programme] to be as diverse as is possible but still meet a standard....relative score of like how good their application is based on all these kind of contextual factors'' (P12). Another requested a technology that helps discover how: ``Close you are to your idealised diversity targets and how close you are to maximising whatever it is you think you're maximising in your performance scores'' (P10).

\section{Part 2: Participatory Design}\label{sec:study2}
\subsection{Prototypes}
\begin{figure}[htbp]
    \centering
    \begin{subfigure}[b]{0.33\textwidth}
        \includegraphics[width=\textwidth]{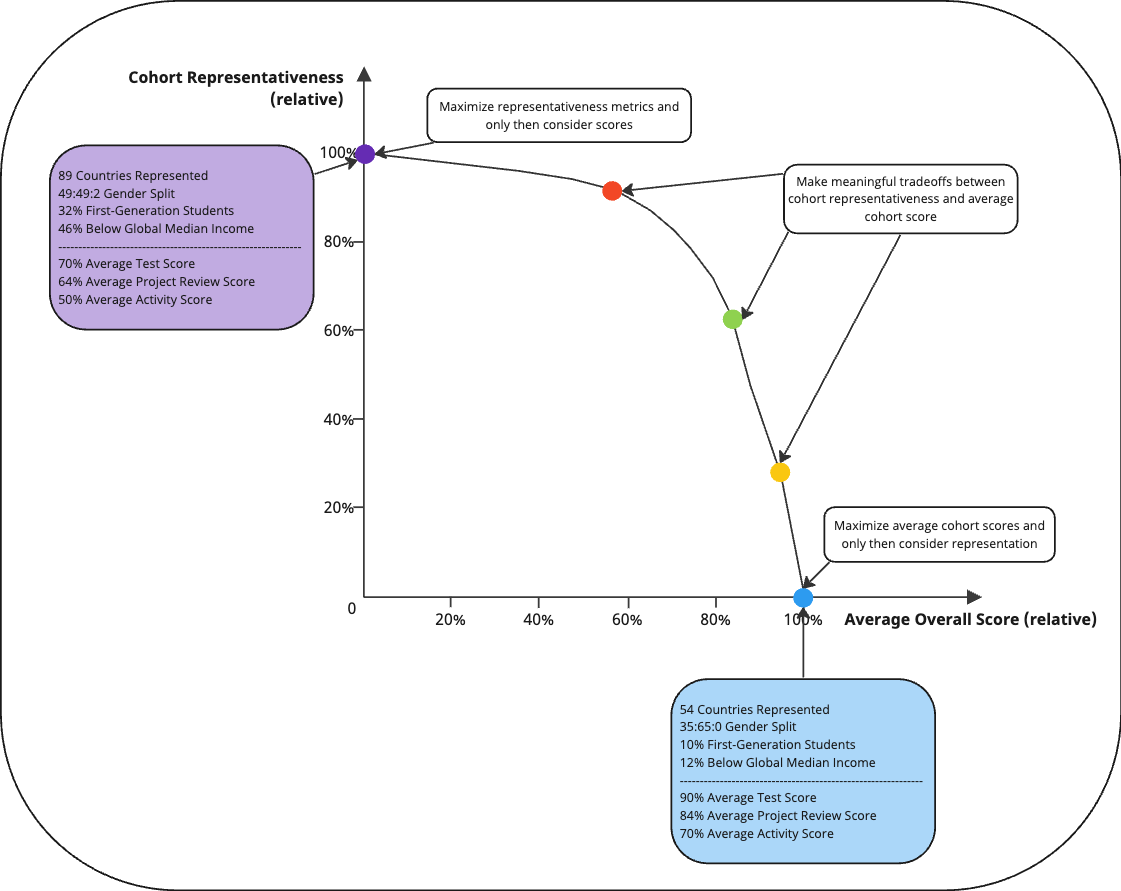}
        \caption{Prototype \ref{fig:representativeness}: Cohort Representativeness}
        \label{fig:representativeness}
    \end{subfigure}
    \hfill
    \begin{subfigure}[b]{0.33\textwidth}
        \includegraphics[width=\textwidth]{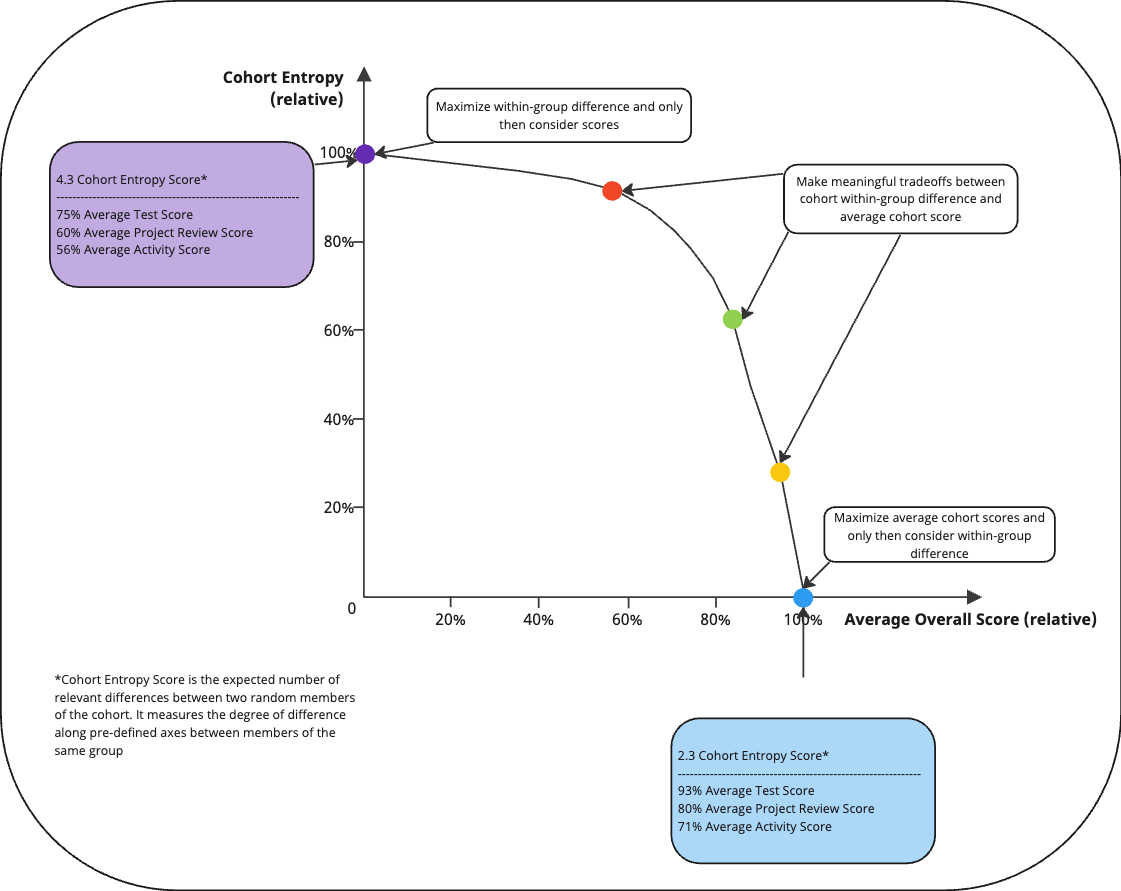}
        \caption{Prototype \ref{fig:entropy}: Cohort Entropy}
        \label{fig:entropy}
    \end{subfigure}
    \hfill
    \begin{subfigure}[b]{0.33\textwidth}
        \includegraphics[width=\textwidth]{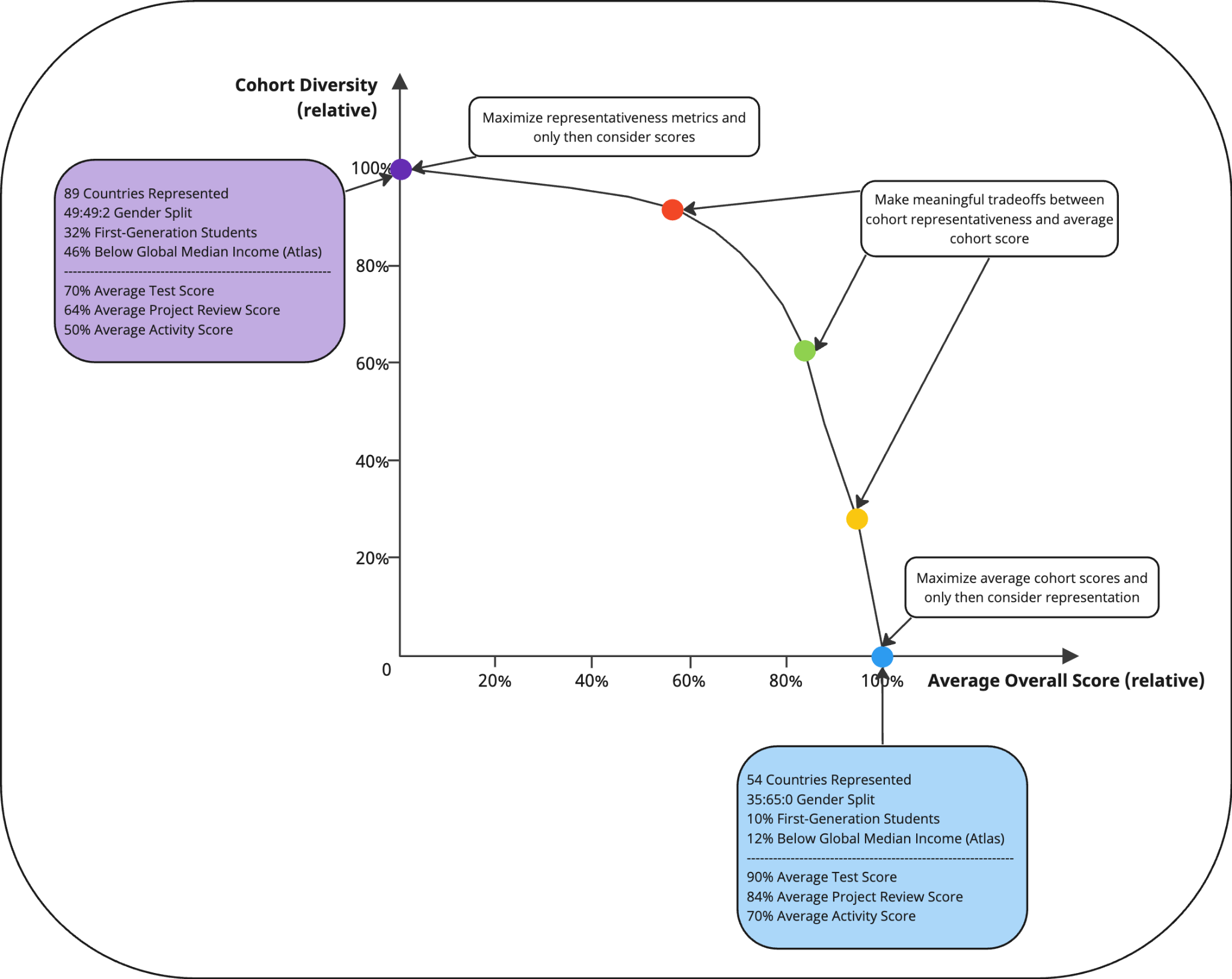}
        \caption{Prototype \ref{fig:diversity}: Cohort Diversity}
        \label{fig:diversity}
    \end{subfigure}

    \medskip

    \begin{subfigure}[b]{0.3\textwidth}
        \includegraphics[width=\textwidth]{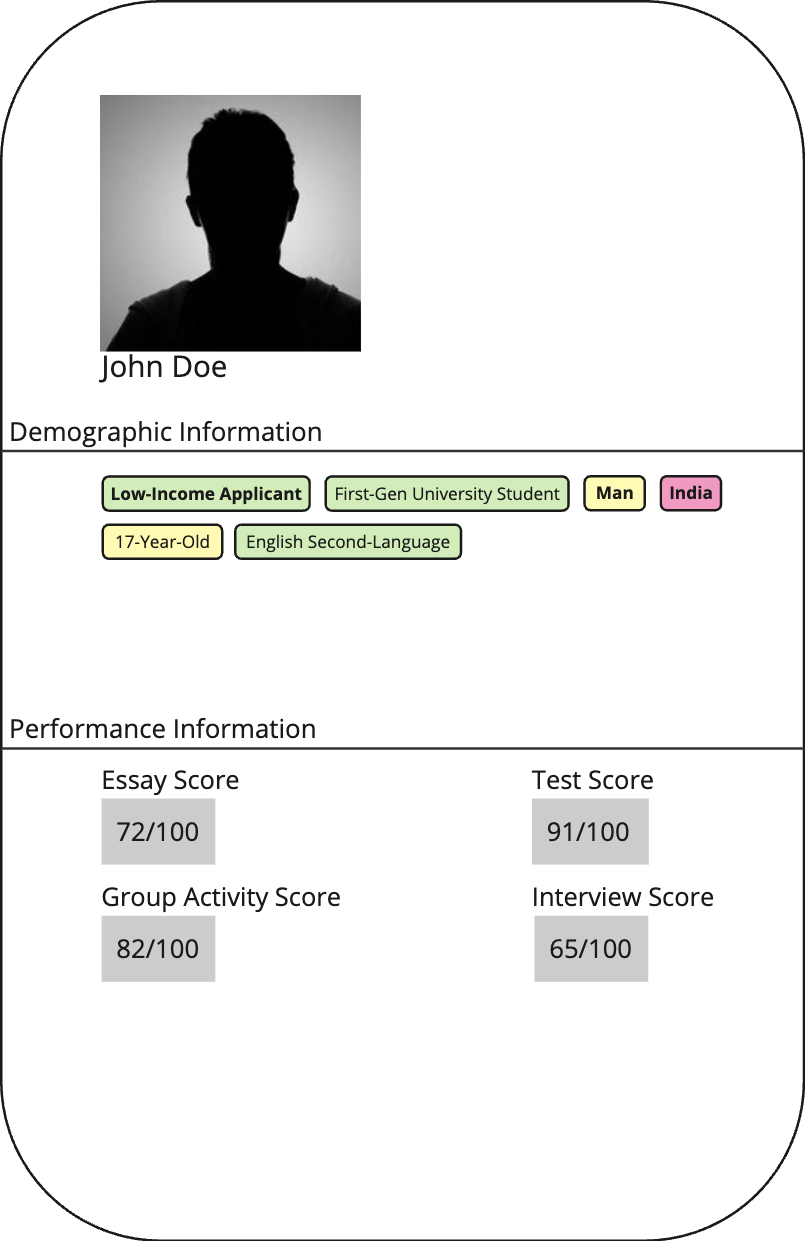}
        \caption{Prototype \ref{fig:demographic}: Applicant Demographic Information}
        \label{fig:demographic}
    \end{subfigure}
    \hfill
    \begin{subfigure}[b]{0.3\textwidth}
        \includegraphics[width=\textwidth]{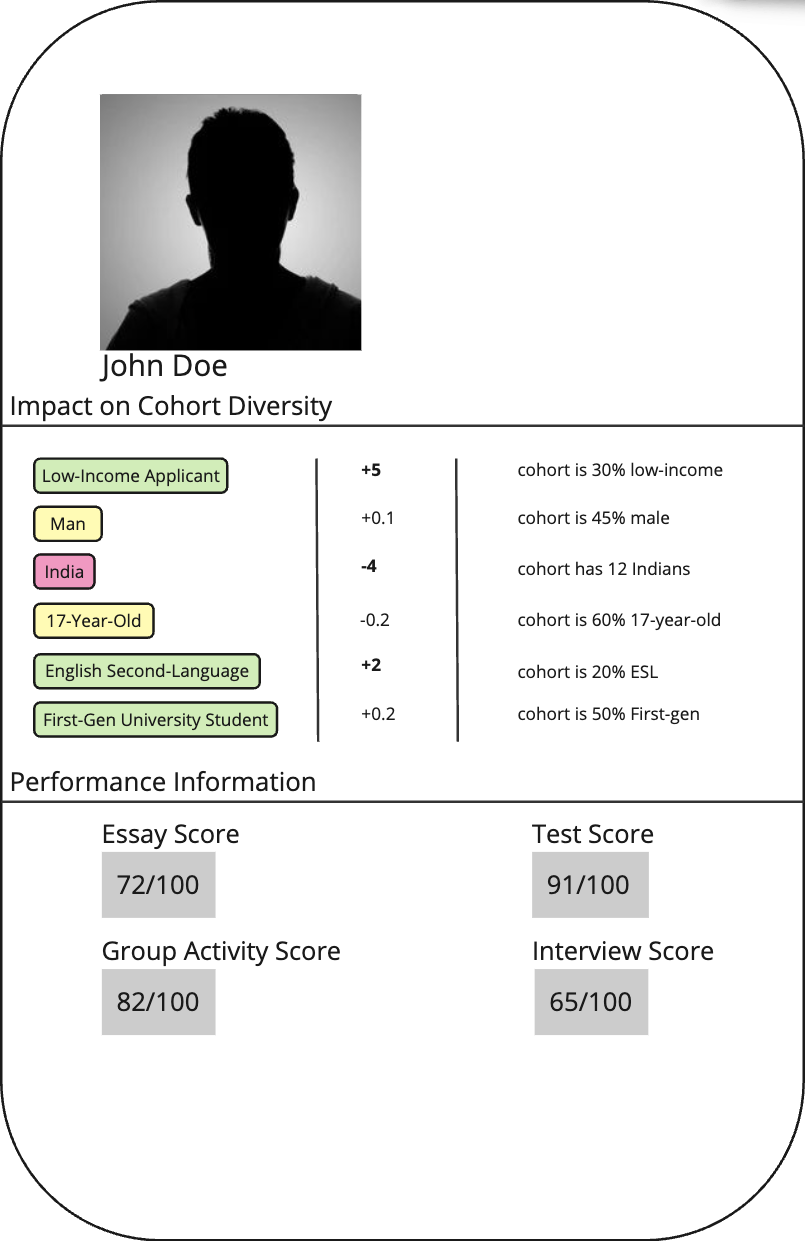}
        \caption{Prototype \ref{fig:impact}: Applicant Demographic Impact on Cohort}
        \label{fig:impact}
    \end{subfigure}
    \hfill
    \begin{subfigure}[b]{0.3\textwidth}
        \includegraphics[width=\textwidth]{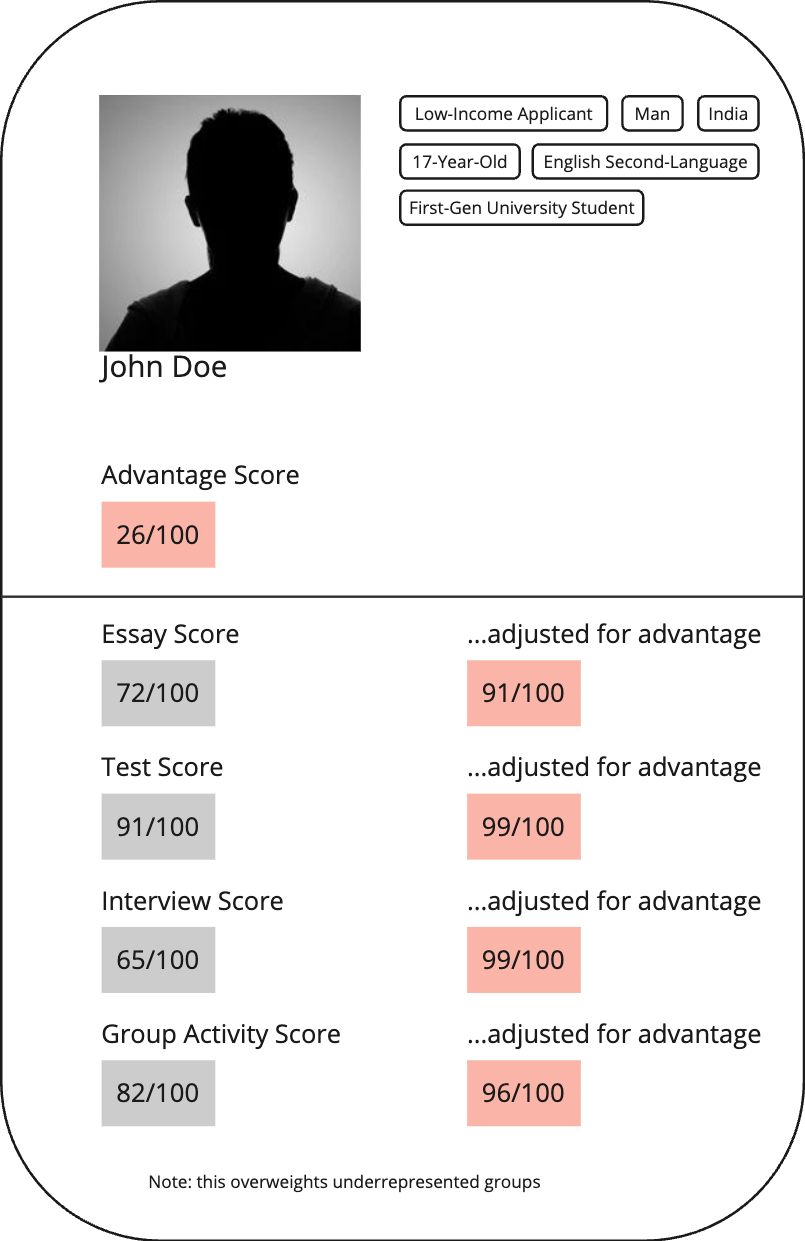}
        \caption{Prototype \ref{fig:advantage}: Applicant Advantage Scores}
        \label{fig:advantage}
    \end{subfigure}
    \caption{These figures depict the prototypes designed based on themes from Section \ref{sec:study1} and used in our participatory design workshops. They are reproduced at a larger scale in Appendix \ref{app:figures}}
    \label{fig:prototypes}
\end{figure}

As a next step, we applied the results of our thematic analysis to design six prototypes following methodology from \textcite{Buchenau_Suri_2000}. These technologies aimed to help selection practitioners better understand and operationalise diversity in their selection processes. We then present these prototypes to participants in participatory design workshops. These prototypes are shown in Figure \ref{fig:prototypes}.

Three of these prototypes (Figures \ref{fig:representativeness}, \ref{fig:entropy}, and \ref{fig:diversity}) present information about the range of possible cohorts participants must choose between, while the other three (Figures \ref{fig:demographic}, \ref{fig:impact}, and \ref{fig:advantage}) present information about an individual applicant relative to a given cohort and a given pool. Furthermore, five of the six prototypes were designed to satisfy definitions of diversity uncovered in Section \ref{sec:study1}.

Figures \ref{fig:demographic} and \ref{fig:impact} are based on the `representativeness' theme, while Figure \ref{fig:advantage} is based on the `contextualising applications' theme. Similarly, Figures \ref{fig:representativeness} and \ref{fig:entropy} draw a distinction in their measurements of diversity, based on the `representativeness' and the `different perspectives' themes, respectively. Figure \ref{fig:entropy}, in particular, defines and employs `entropy' as a metric. This reflects that, when aiming to get different perspectives in the same room, the goal is not to represent any target population; rather, we desire that everyone in our group be as different from the remainder of the group as possible.

\subsection{Workshop Methodology}\label{ssec:methods2}
As the participants all come from two separate talent investment programs, we run one workshop for each group \cite{Buchenau_Suri_2000}. Before presenting to the broad audience of each group, we submit our figures to one primary contact (also a participant in the study) at each organisation, then run informal, 15-minute `pilot' one-on-one workshops with this primary contact. Here, we primarily sought approval to use these figures in workshop with the broader team of practitioners, but we also collected minor feedback and tweaked the prototypes based on this feedback. In one organisation's pilot workshop, the primary contact requested Figures \ref{fig:representativeness} and \ref{fig:entropy} be combined into one prototype with `Diversity' as the Y-axis, as the organisation already has an internal working definition of diversity (that aligns very closely with what we mean by representativeness). This can be seen in Figure \ref{fig:diversity}. Participant grouping for the workshops can be seen in Table \ref{tab:participants}; the results for this analysis are attributed by group, rather than individual, to reflect the cooperative nature of the task.

Our central research questions for this workshop are:

\begin{enumerate}
    \item What prototypes best promote diversity?
    \item What elements of these prototypes facilitate their success?
\end{enumerate}

\noindent Or, for each prototype: ``How and why does this prototype promote diversity?''. In each workshop, we ask participants to consider each prototype in turn, and to discuss how they might use it in their selection process. We then ask participants to consider how these prototypes might fit into their current selection process, and how they might change their process to better incorporate these prototypes. Finally, we ask participants to consider how their current selection process might make best use of these prototypes, and whether they think these prototypes would be beneficial. 

Following the methodology of \textcite{Gatian_1994,Griffiths_Johnson_Hartley_2007}, at the end of each workshop, we ask participants to highlight their favourite prototype.

A question-by-question protocol for these workshops can be found in Appendix \ref{app:protocol2}.

\subsection{Results}\label{ssec:results2}
\subsubsection{Participants Preferred Prototype \ref{fig:impact}}
As part of the workshop, participants were asked to mark their favourite prototypes \cite{Gatian_1994,Griffiths_Johnson_Hartley_2007}. These favourites have been collated in Table \ref{tab:favourites}, and it can be seen here that Prototype \ref{fig:impact} was by-far the favourite in both groups.

\begin{table}[htbp]
    \centering
    \caption{This table tallies the number of participants indicated that a given prototype was their favourite. The overwhelming favourite was prototype 5, which shows an individual applicant's impact on the cohort.}
    \label{tab:favourites}
    \begin{tabularx}{.25\textwidth}{lX}
        \toprule
        \textbf{Prototype} & \textbf{Favourites} \\
        \midrule
        Prototype \ref{fig:representativeness} & 1 \\
        Prototype \ref{fig:entropy} & 1 \\
        Prototype \ref{fig:diversity} & 0 \\
        Prototype \ref{fig:demographic} & 1 \\
        Prototype \ref{fig:impact} & 10 \\
        Prototype \ref{fig:advantage} & 0 \\
        \bottomrule
    \end{tabularx}
\end{table}

\subsubsection{Both Groups Rely on Idiosyncratic Notions in Selection}
When participants were placed in a practical selection scenario and shown technology prototype information about (hypothetical) applicants, they compared applicants to idiosyncratic profiles that they desired.

When evaluating Prototype \ref{fig:advantage}, G1 used advantage scores to seek out: ``Diamonds in the rough'' (G1), i.e., talented applicants from disadvantaged backgrounds who lacked the polish of their more privileged counterparts.

Prototype \ref{fig:entropy} was initially confusing to G2, as the `entropy' definition used was unfamiliar to the participants: ``Entropy is chaos in chemistry. How does this relate to our usage here?'' (G2). Thus, when interacting with Prototype \ref{fig:entropy}, G2 understood `entropy' to be variety in cognitive skill-set and personality type. In particular, the group sought to ensure that the chosen cohort contain `glue' people, who improve overall cohort cohesion: ``For people working together, it's useful to have someone who is that `glue''' (G2). 

\subsubsection{Organisations and Participants Are Interested in Different Diversities}
When presented with Prototypes \ref{fig:representativeness} and \ref{fig:entropy}, G2 was given the demonstrative definitions of `representativeness' and `entropy' (visible on the prototypes). G2 quickly understood the differences between the figures to be that Prototype \ref{fig:representativeness} sought to support considerations of demographic representativeness, while Prototype \ref{fig:entropy} sought to support placing different perspectives (be they cognitive skill-sets or personality types) in the same room. Participants were then interested to know the relationship between these prototypes in practice: ``If we maximise based on [Prototype \ref{fig:representativeness}] scores, what would the Entropy scores be?'' (G2).

However, though individual participants took great interest in Prototype \ref{fig:representativeness}, G2 ultimately acknowledged that the program was most interested in building a cohort from people with different perspectives to facilitate collaboration: ``For [our] cohorts, [we] want a balance [of personality types] and [we] want them to be collaborative'' (G2). At the same time: ``Let's track but not use [program-specific measure of representativeness]'' (G2).

\subsubsection{Different Tools are Useful at Different Application Stages}
Participants in both groups variously expressed an anxiety about measuring the relevant dimensions. ``What are our metrics and are they reliable?'' (G2) was echoed by several G2 participants. ``If it's all self-report, then we can't do anything with it'' (G1). 

However, both organisations noted that their selection processes involve a variety of stages. Different tools are useful in different stages. When speaking of Prototype \ref{fig:entropy}, one participant said: ``This is better post-interview than it is pre-interview'' (G2), as interviews will collect observational data on many of these characteristics. I.e., while certain tools may rely on measurements that cannot be collected until later stages, others will be more useful earlier in the selection process.

Similarly, different output modes were useful at different stages. Note that Prototypes \ref{fig:demographic} and \ref{fig:impact} contain similar information, but that Prototype \ref{fig:impact} displays this information in greater detail. When reviewing both prototypes, G1 found Prototype \ref{fig:demographic} preferable in the earlier stages of decision-making, while Prototype \ref{fig:impact} had the greatest utility later. ``[We] can't send [Prototype \ref{fig:impact}] as a pre-read, but [Prototype \ref{fig:demographic}] makes more sense in isolation, so better for a pre-read'' (G1). On Prototype \ref{fig:demographic} in particular, one participant noted: ``Helpful for the process, not so much for the [final cohort selection]'' (G1). Another said of Prototype \ref{fig:impact}: ``This has the most potential at the later stages of decision-making'' (G1).

At the other extreme, while the cohort-level tools did not spark discussions about particular individuals, they did spark earlier-stage, higher-level discussions. Most obviously, both programs discussed specific tradeoffs between measured individual applicant aptitude and cohort diversity: ``Real decision-making always sees tradeoffs like these.... This chart helps you figure out the level of compromise you're willing to make on both axes'' (G2); ``[It] makes sense that top scoring candidates don't necessarily help you build the most diverse cohort.... 5-10\% [of the cohort] really get to a struggle between quality and diversity.... If this chart were real (rather than hypothetical), and you could see who you were losing, this would be useful'' (G1). (In the hypothetical,  participants from G1 settled closer to the centre of the frontier: ``[Our] target here is to look somewhere [from] red to yellow'' (G1). However, they also noted that ``Some candidates get a big diversity boost and score terribly'' (G1).)

In one case, participants also discussed broader, program-level concerns. Participants spent time debating ``Is the program needs-based or merit-based?'' (G1). They noted  that ``This chart helps [facilitate that discussion]'' (G1), but did not ultimately conclude one way or the other.

\subsubsection{The Right Balance of Quantitative and Qualitative Information is Key}
Participants simultaneously expressed gratitude that the prototypes were as simplified as they were, and a desire for more detail. One participant from G1 noted that the prototypes are: ``Very constrained in terms of what is being shown.... This doesn't include all of the factors, but for the decision that's good, because it prevents info overload'' (G1). However, participants G2 frequently requested additional quantitative information. In the case of the Prototype \ref{fig:impact} (participants' favourite prototype, as can be seen in Table \ref{tab:favourites}), participants requested a number of other metrics: ``Advantage score'' (G2), ``Impact on entropy'' (G2), ``[A summary of] these scores together'' (G2), and ``A composite impact on cohort diversity'' (G2).

Meanwhile, participants from G1 requested qualitative information: ``We need to know more about the applicants' backgrounds'' (G1). Other requested information included: ``Comments from selectors'' (G1) and ``A narrative summary...written by the selector team'' (G1). 

This suggests a discrepancy between the two groups' preference for the balance between qualitative and quantitative information. While G2 expressed a desire for unifying quantitative metrics: ``A single score for disadvantage [or] need, while recognising its flaws, could provide one read of an individual's circumstances'' (G2), G1 expressed a desire for individual, qualitative information: ``We're using quantitative to sift through qualitative.... [We] need to include comments from selectors'' (G1).

\section{Design Recommendations}
\begin{figure}[htbp]
    \centering
    \includegraphics[width=0.9\linewidth]{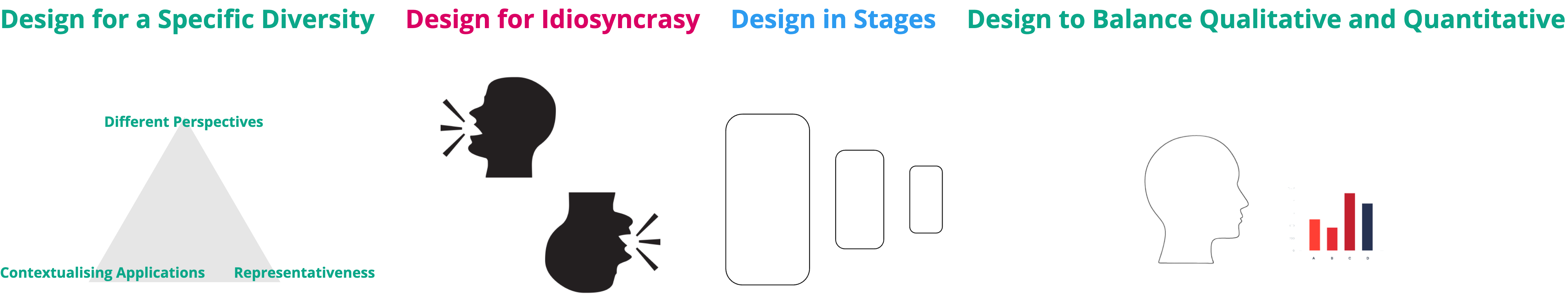}
    \caption{This figure illustrates our four key design recommendations to others building tools to support the selection of diverse talent, i.e., `design...': `...for a specific diversity', `...for idiosyncrasy', `...in stages', and `...to balance qualitative and quantitative'.}
    \label{fig:recommendations}
\end{figure}

\subsection{Design for a Specific Diversity}
In the initial interviews, participants were often vague when asked to define diversity. However, when asked to expand on why diversity is important, or on what dimensions of diversity they prioritised, it became clear that `diversity' included three separate (and sometimes competing) definitions. We have termed these: `representativeness', `different perspectives', and `contextualising applications'. When designing tools to assist a target organisation in considering diversity, we suggest designers first clarify through human-centric methods which definitions of diversity the target organisation seeks to consider, then designs to support those specific definitions. That is, designers should follow \textcite{VanKleek_Seymour_Binns_Shadbolt_2018}'s paradigm of `respectful' design and build technology to best serve the needs of the selection practitioners who will use it.

\subsection{Design for Idiosyncrasy}
One key note revealed through this process is that different decision-making processes have philosophical underpinnings, desiderata, and anecdotal definitions that impact their selections. These idiosyncrasies should be discovered early in development and designed for in any technical solution. We suggest participatory design as a mechanism for achieving this. We observed a strong relationship between participant feedback in interviews and their satisfaction with the prototypes.

For example, both talent investment programs we worked with have created specific personas they look for. For example, one group discussed `glue' people who helped groups function cohesively. The other group discussed `diamonds in the rough', talented youth systemically undervalued due to their backgrounds. Where these aspects were included, participants showed strong interest in the prototypes. Where they were excluded, participants often asked for these to be added.

\subsection{Design in Stages}
Much of what participants desired at one stage of decision-making was mutually exclusive with that they desired at other stages. For example, G1 desired Prototype \ref{fig:demographic} as a: ``Pre-read'' (G1) due to its simplicity, but preferred the detail of Prototype \ref{fig:impact} ``In the room'' (G2). Thus, it is crucial for designers to consider what stage of decision-making their tools are designed to support, and to design appropriate levels of detail, abstraction, and engagement accordingly.

\subsection{Design to Balance Qualitative and Quantitative}
Participants often noted that the prototypes were missing key qualitative information about applicants. This qualitative information is crucial in holistic considerations of each applicant. However, when allowed to consider only qualitative information, participants obscure tradeoffs they are forced to make between different program goals. In particular, while individual-level goals are often clear, cohort-level goals (such as diversity) are easier to delay or ignore. Thus, without quantitative tools to frame the discussion, participants noted that they were often forced to make cohort-level considerations ad-hoc and towards the end of their decision-making process.

Thus, while qualitative information is crucial, it is also important to present the quantitative information necessary to make these tradeoffs salient. Ultimately, final selection decisions are made by panels of trained selectors, but in the absence of both quantitative and qualitative information to guide these decision-makers, they would be forced to make decisions that are less well-informed than they could be.

\section{Discussion}\label{sec:disc}
\subsection{The Diversity Triangle}
\begin{figure}[htbp]
    \centering
    \includegraphics[width=\textwidth]{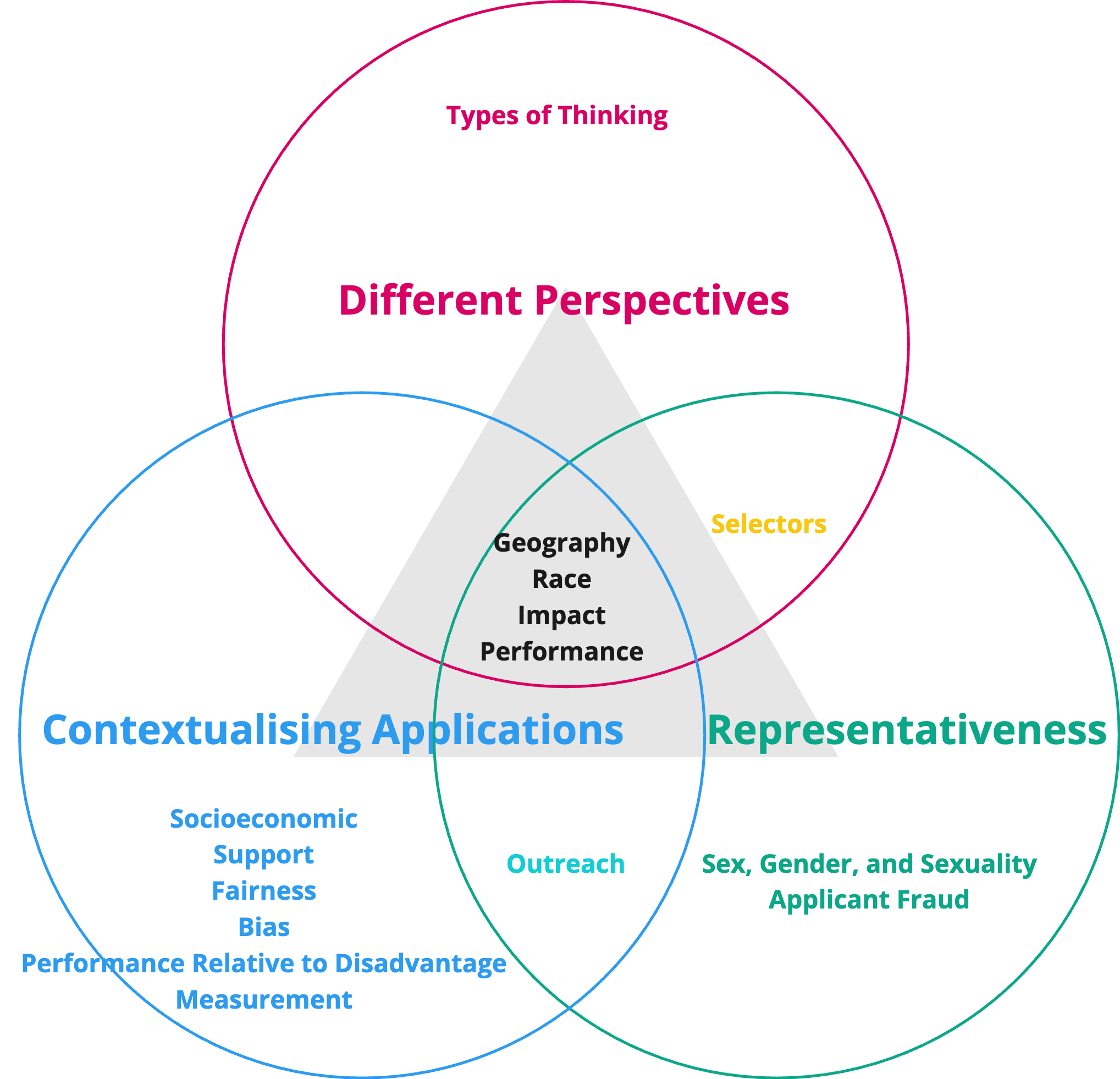}
    \caption{This figure depicts the \emph{Diversity Triangle}, three differing definitions of diversity selection practitioners expressed when discussing a diverse cohort. We also relate the Diversity Triangle to each other theme or subtheme participants mentioned.}
    \label{fig:div_triangle}
    \Description{A Venn diagram with `Diversity' in the centre, featuring `Different Perspectives', `Representativeness', and `Contextualising Applications' on each corner. Each theme has been placed into one of the overlapping sections.}
\end{figure}

In Section \ref{sec:study1}, participants variously identified the word `diversity' with three themes, which we have taken to signify definitions of diversity: `representativeness', `different perspectives', and `contextualising applications'. As these three definitions are central to our research questions, we privileged them over Section \ref{sec:study1}'s other themes in Section \ref{sec:study2}, where we engaged in scenario speed dating and experience prototyping designed to satisfy different definitions of diversity. In this section, we map the themes to the three definitions in the \emph{Diversity Triangle} (Figure \ref{fig:div_triangle}).

\subsection{The Relationship Between the Diversity Triangle and Theories of Change}
\subsubsection{Representativeness}
Each of the definitions from the Diversity Triangle (Figure \ref{fig:div_triangle}) implies different program values and a different theory of change. The representativeness
definition relies on social theories by which diversity is intrinsically valuable. Several participants seemed to believe representativeness here was intrinsically valuable, which aligns with the \textcite{morris1984origins} account under which contemporary diversity norms emerged from loci of oppression affecting underrepresented groups. However, some participants also noted instrumental reasons to value representativeness that align closely with \textcite{peters2021hidden,page_difference_2007}'s argument that people from different backgrounds often posses unique and germane knowledge. One participant noted that a team can: ``Better serve a community if they represent [that community]'' (P9). Another theory from \textcite{friedler_impossibility_2016} discusses measurement bias. Notably, if we assume that talent is equally distributed across some partition, then the most talented cohort should also be representative. However, \textcite{friedler_impossibility_2016} note that we often observe in practice that performance is not equally distributed across these partitions. Depending on one's worldview, one might see this as due to inherent differences between those groups, or due to structural bias that causes differences in construct observability between them \cite{friedler_impossibility_2016}. In the latter case, representativeness would be a proxy for fairness. Thus, representativeness is pro-social: society is better served when resources are distributed to a representative group of people.

\subsubsection{Different Perspectives}
The argument for placing different perspectives in the same room is often instrumental. While `diversity' on the whole is often spoken of as an intrinsically valuable broader benefit, the aim of placing different people in the same room is often only to benefit the people in that room. \textcite{page_diversity_2010} argues that `cognitively' diverse groups outperform homogeneous groups on some tasks; some participants similarly contend that it improves cohort-level task performance. Other participants echo \textcite{wylie2006introduction}'s argument that it allows participants to better learn from each other. In either case, the benefit is primarily organisational rather than social.

\subsubsection{Contextualising Applications}
The argument for contextualising applications is twofold. Most often, participants make a systemic critique here. That is, the world is incredibly unjust, and we want to distribute resources differently, but in a talent selection process we still have to operate in the unequal world. Thus, in order to correct for that injustice, we must give more resources to those who have less. This could be seen as a form of distributive justice or `affirmative action'. Participants argue (perhaps relatedly) that appropriately contextualising applications results in more successful applications from disadvantaged groups; this, in turn, allows these applicants to have a positive impact on their groups. Relatedly, either due to measurement bias, or due to differences in performance brought on by disparate access to resources, support, and opportunities, we may find that applicants from disadvantaged backgrounds appear worse on paper. In either case, correcting this through contextualisation may also build a more fair selection process and a more just world, yielding great benefit to society. On the organisational side, if contextualising applicants allows for the admission of more applicants from marginalised groups, they may be better suited to critiques of existing power structures, as they may have more informative experiences of oppression \cite{mills2015blackness}.

\subsection{A Cautionary Note On Designing for Diversity}
Decisions such as scholarship selection have long-reaching impacts for the applicants. A successful applicant to a scholarship program might thus attend a university they could not have otherwise attended. There, they will acquire skills and a network that will continue to impact them later in life. These impacts may even affect those close to them, as their better circumstances likely better the circumstances of their communities. Thus, it is simultaneously important to ensure that these opportunities are dispersed fairly and to ensure that all demographics are included among those selected.

This increases the importance of selecting a diverse group of people, and mandates that we do all we can to do so. Quantitative decision support tools may have a role to play in this, but two problems remain: data collection and data processing. Primarily, diversity data is often self-reported. Thus, we cannot use this naively to generate diverse cohorts, as doing so would yield a competitive advantage to candidates who lie on their declarations. Secondly, processing data automatically has its own drawbacks. AI systems are prone to bias \cite{friedler_impossibility_2016}. And as these systems improve, it is unclear if they will help or harm our ability to select diverse cohorts.

However, it is important to note that both the problems of data collection and processing are not unique to this workflow. In fact, algorithmically-supported diversity considerations may correct for data collection and processing issues elsewhere. Bias is a major consideration in data collection \cite{friedler_impossibility_2016}; heterogeneous biases in measurements of talent may be corrected by ensuring diversity across these metrics: ``Assessments are usually biased measures of what we actually care about, and that opportunity often correlates with positive error terms in assessments' measure of underlying skills.... [Diversity considerations] correct for the inadvertent affirmative action against underprivileged individuals implicit in using biased assessments'' (P1). Similarly, much like AI systems, human selectors are prone to their own heterogeneous biases; in some sense, `design for idiosyncrasies' reflects the notion that decision support systems must at times help recognise and mitigate these biases.

\section{Limitations and Future Work}
In Studies 1 and 2 we build tools with which participants report satisfaction, but increasing participant satisfaction does not necessarily improve decision-making. In particular, technology that makes difficult decisions less painful may be well-received, while technology that makes these decisions more salient may be less popular but more impactful \cite{Lipton,Miller_2023}. Our themes speak to participant perceptions of diversity and our design recommendations speak to participant desiderata from support tools. We assume throughout that, in solving for these considerations, we can help organisations select better cohorts. However, we may find that these tools fail to improve decision-making with regards to diversity in practice. Future work should investigate this possibility through implementation of our prototypes in field settings.

\textcite{Venn-Wycherley_Kharrufa_Lechelt_Nicholson_Howland_Almjally_Trory_Sarangapani_2024} contend that HCI literature focused on educational contexts should consider both educator and student. Two distinctions distance our work from theirs: first, selection is distinct from pedagogy in that our decision subjects are not necessarily beneficiaries (while students are beneficiaries of their institution, applicants do not become beneficiaries unless they are selected); second, scholarships are distinct from educational institutions in that scholarship program benefits primarily focus on assisting beneficiaries in accessing educational institutions, while educational institutions primarily educate beneficiaries. Both distinctions create distance between selectors and applicants beyond that between teachers and students and pose challenges in engaging decision subjects. Nonetheless, future work should engage scholarship applicants to understand their definitions, stances, and considerations with respect to diversity. 

While not a limitation, we have intentionally set aside themes such as `outreach' and `selectors' that relate more to other aspects of selection processes than to the act of selection itself. We hope future work will consider these facets of selection programs, especially when designing tools to support thinking around diversity.

\section{Conclusion}
This research answers the crucial question: ``What is diversity?'', from the perspective of scholarship and talent investment selection practitioners. In doing so, we illuminate the multifaceted nature of diversity in scholarship selection and emphasise the critical need for tools that support considerations of diversity in decision-making. Our findings reveal that achieving true diversity involves navigating the complex interplay between three occasionally conflicting definitions: representativeness, diverse perspectives, and the contextualisation of applicants' backgrounds. By engaging in participatory design, we build six prototypes with our practitioners; this process reveals four design recommendations: design for a specific diversity, design for idiosyncrasy, design in stages, and design to balance quantitative and qualitative. This work demonstrates that, when thoughtfully designed, technology can empower selection processes to be more equitable, inclusive, and transparent. The broader implication is that such advancements have the potential to reshape how diversity is operationalised, ensuring that it is both a measurable outcome and a core value in shaping the future of talent identification. As we move forward, the integration of these tools into real-world practices will be pivotal in fostering truly diverse and representative groups in global scholarship programmes and beyond.

\printbibliography

\appendix

\section{Individual Interview Protocol}\label{app:protocol1}
For the individual interviews, we follow a semi-structured protocol. Following the methodology of \textcite{braun_using_2006}, we do not limit our analysis to these questions. Instead, we deviate from this script as guided by the conversations and our overarching research questions, then we allow themes to emerge naturally from the data. Our interview research questions (also found in Section \ref{ssec:methods1}) are:

\begin{enumerate}
    \item What is diversity?
    \item Which elements of diversity matter in a selection context? Why?
    \item How could technology assist in operationalising diversity?
\end{enumerate}

We interview 15 individuals from two different talent identification organisations. We conduct each interview separately. We first ask a few questions about the factors that go into decision-making:

\begin{enumerate}
    \item We're going to take a step back and discuss a hypothetical selection scenario for a fellowship for a group of young people. In this scenario, you have full control over who is selected.
    \item Could you please list some things you think are important in deciding who to accept?
    \item (Can skip) Which of (these things) are about the individual applicant's performance?
    \item What are (these remaining things) about? 
    \item (Or:) Why are (these things) important?
\end{enumerate}

We then ask participants to define diversity, to break it down into elements, and to discuss why diversity is important:

\begin{enumerate}
    \item Now I want to talk about diversity. Keeping your list in mind, can you please define diversity?
    \item (If the definition is too short) Could you please elaborate on (pick a part)
    \item Why do we care about this definition of diversity?
    \item Now, if you were going to break your definition into some elements or facets, what would those be?
    \item (If they talk about holistic diversity) What considerations are important when looking at diversity holistically?
    \item (If elements are vague) How does (pick a metric) factor into your understanding of (element)?
    \item Which elements or considerations are most important?
    \item How do we measure these facets of diversity?
    \item (If this measurement isn't concrete) Imagine you had a ``magic metric'' that perfectly measured diversity. What does this metric do?
\end{enumerate}

Next, we run two short exercises from the participatory design literature. The first is called ``crazy 8s'', wherein participants are given 8 minutes to come up with 8 ideas. For these ideas, we ask participants to think about technologies that might help them better understand diversity in selection:

\begin{enumerate}
    \item Now I want to talk about technology we can build to support thinking about tradeoffs around diversity. Remember that we're stepping away from existing processes and solutions.
    \item We're going to start with an exercise called ``crazy 8s''. For the next 8 minutes, we're going to spend one minute each developing a technology that might help us better understand diversity in selection. These technologies don't have to make sense or be possible; I just want you to think of things that might help you think through diversity. This activity is difficult; don't worry if you find yourself struggling or sounding silly.
    \item Take a second to think about a technology. When you're ready, please describe it.
    \item (If the technology is unclear) Could you please elaborate on (unclear part)?
\end{enumerate}

The second is called ``the magic app'', wherein participants are asked to elaborate on a single idea for an application, waving away technical details as `magic':

\begin{enumerate}
    \item Now let's dig deeper into one hypothetical ``magic app'' designed to help us better understand tradeoffs around diversity in selection. The magic app can do anything you might want it to in any way you might want. What does your magic app do?
    \item (If the app has visuals) What do your visuals look like?
    \item (If the app is pure text) What sorts of visualisations might help you?
    \item (If the app has buttons or sliders) What do your buttons do?
    \item (If the app doesn't have any interactivity) How might you interact with this app?
    \item Now we are going to split the app out into different ``pages''
    \item (If they haven't already done this) The individual-level page: each applicant will have their own individual-level page, which will say things about that applicant
    \item (If they haven't already done this) The cohort-level page: each possible cohort will have its own cohort-level page, which will update any time we make changes to the cohort.
    \item (For each page) What happens on this page?
    \item What is the experience of using this page like?
    \item (If the page has visuals) What do your visuals look like?
    \item (If the page is pure text) What sorts of visualisations might help you?
    \item (If the page has buttons or sliders) What do your buttons do?
    \item (If the page doesn't have any interactivity) How might you interact with this app?
    \item (For each different feature of the page) What makes (feature) useful to you?
    \item Thank you! Is there anything else you would like to add?
\end{enumerate}

\section{Participatory Design Workshop Protocol}\label{app:protocol2}
For the participatory design workshops, we split our participants by organisation. As some individuals could not attend the second session, we have one group of 6 and another of 7. As these are much larger group discussions, we deviate further from our protocol.

The task for these workshops consists of hands-on sessions with different technologies. The technologies are designed and mocked up based on the thematic analysis of the interviews. These technologies are presented to participants via Miro, where they are free to interact with and annotate them. Our research questions for this workshop are:

\begin{enumerate}
    \item What prototypes best promote diversity?
    \item What elements of these prototypes facilitate their success?
\end{enumerate}

Or, for each prototype: ``How and why does this prototype promote diversity in talent identification?''. Again, though we write a list of questions targeted at these questions, we do not limit our analysis to these questions \cite{braun_using_2006}. Instead, we deviate from this script as guided by the conversations and our overarching research questions, then we allow themes to emerge naturally from the data \cite{braun_using_2006}. For each technology prototype shown, we have questions:

\begin{enumerate}
    \item This prototype describes... Are any of you familiar with this?
    \item In what follows, we're going to discuss this prototype. Let's start with: is it easy to read for you? What does it say? 
    \item What questions do you have upon seeing this prototype? Feel free to write these down.
    \item How would you use this prototype in a hypothetical selection procedure?
    \item How (else) would this prototype fit into your current selection procedure?
    \item How would your current selection procedure make best use of this prototype? Would the process need to be changed? Do you think this would be beneficial?
\end{enumerate}

Finally, at the end of our workshop, after we have covered all of the prototypes, we ask participants to place a star next to their favourite prototype on the Miro board \cite{Gatian_1994,Griffiths_Johnson_Hartley_2007}.

\section{Images and Descriptions of Prototypes}\label{app:figures}
This appendix contains figures depicting the prototypes used in Section \ref{sec:study2}'s participatory design workshops at a larger scale.

\begin{figure}[htbp]
    \centering
    \includegraphics[width=0.9\linewidth]{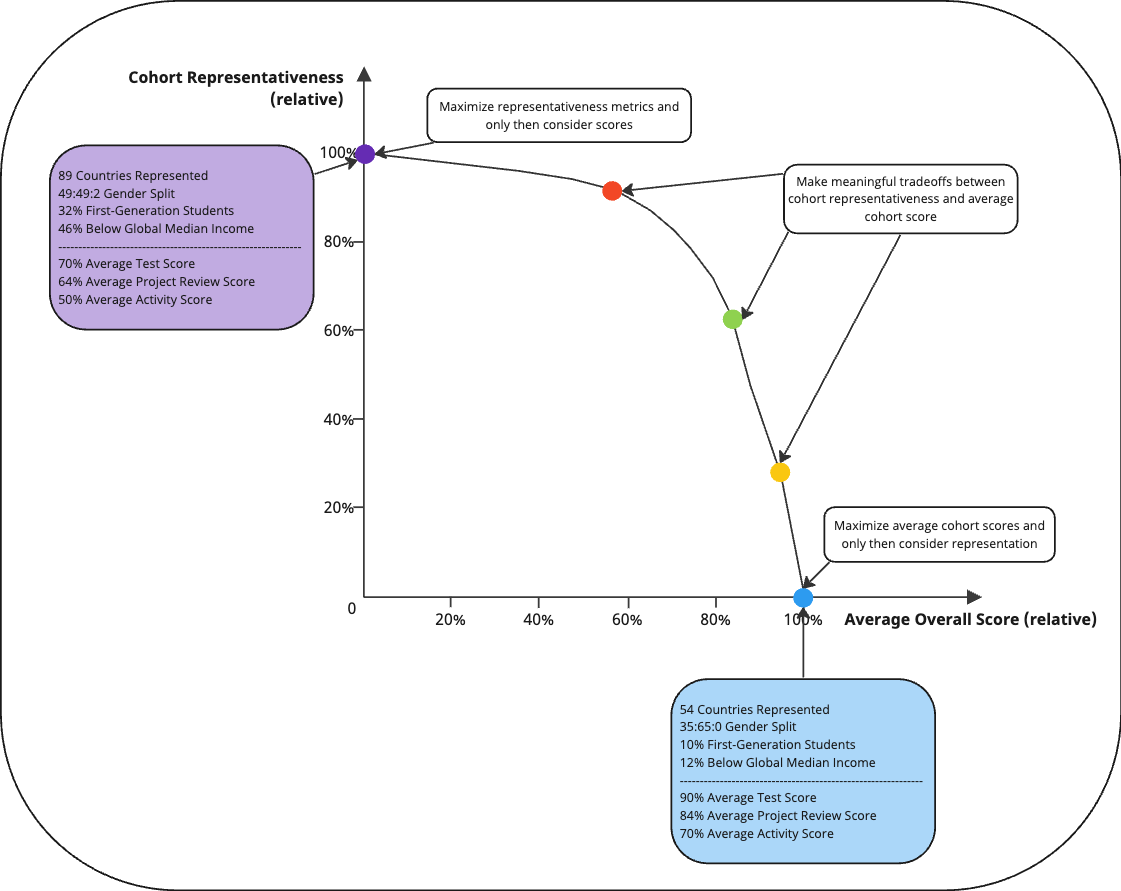}
    \caption{This figure reproduces Prototype \ref{fig:representativeness} at a larger scale.}
    \label{fig:representativeness_full}
\end{figure}

\begin{figure}[htbp]
    \centering
    \includegraphics[width=0.9\linewidth]{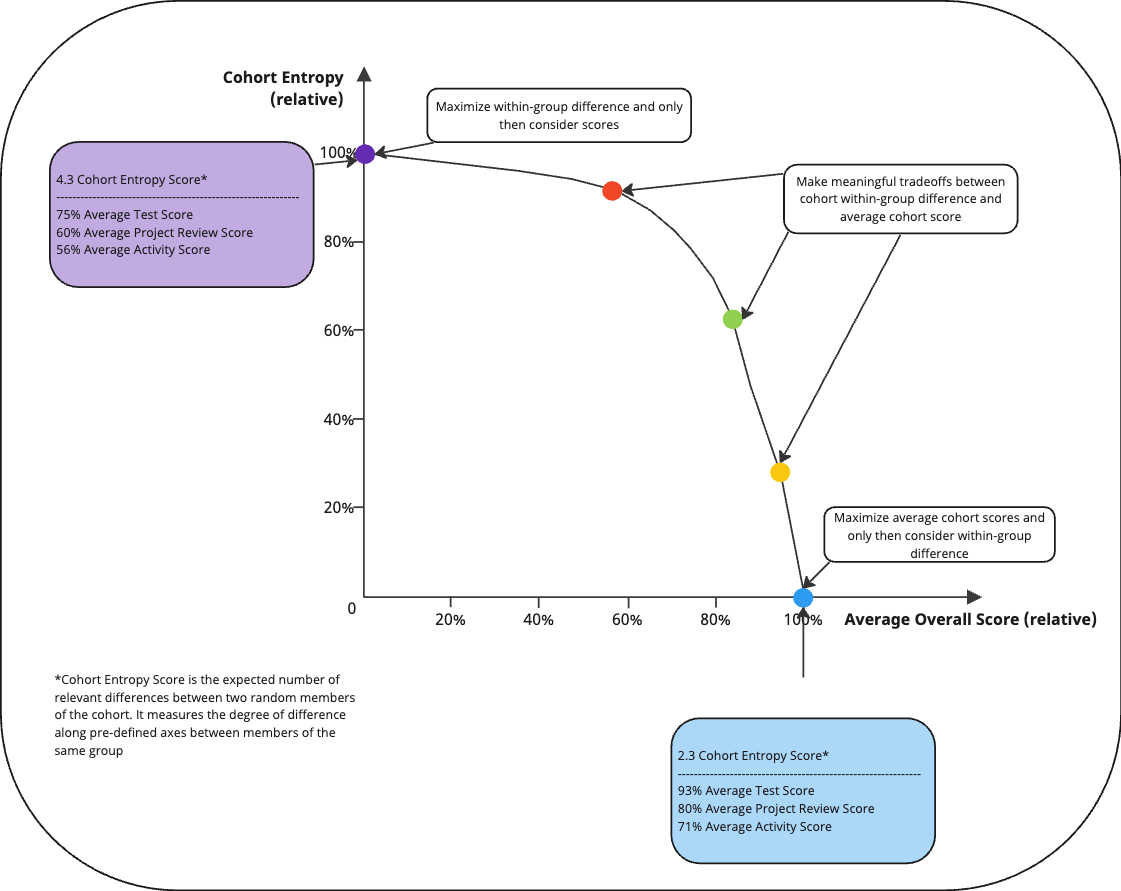}
    \caption{This figure reproduces Prototype \ref{fig:entropy} at a larger scale.}
    \label{fig:entropy_full}
\end{figure}

\begin{figure}[hbtp]
    \centering
    \includegraphics[width=0.9\linewidth]{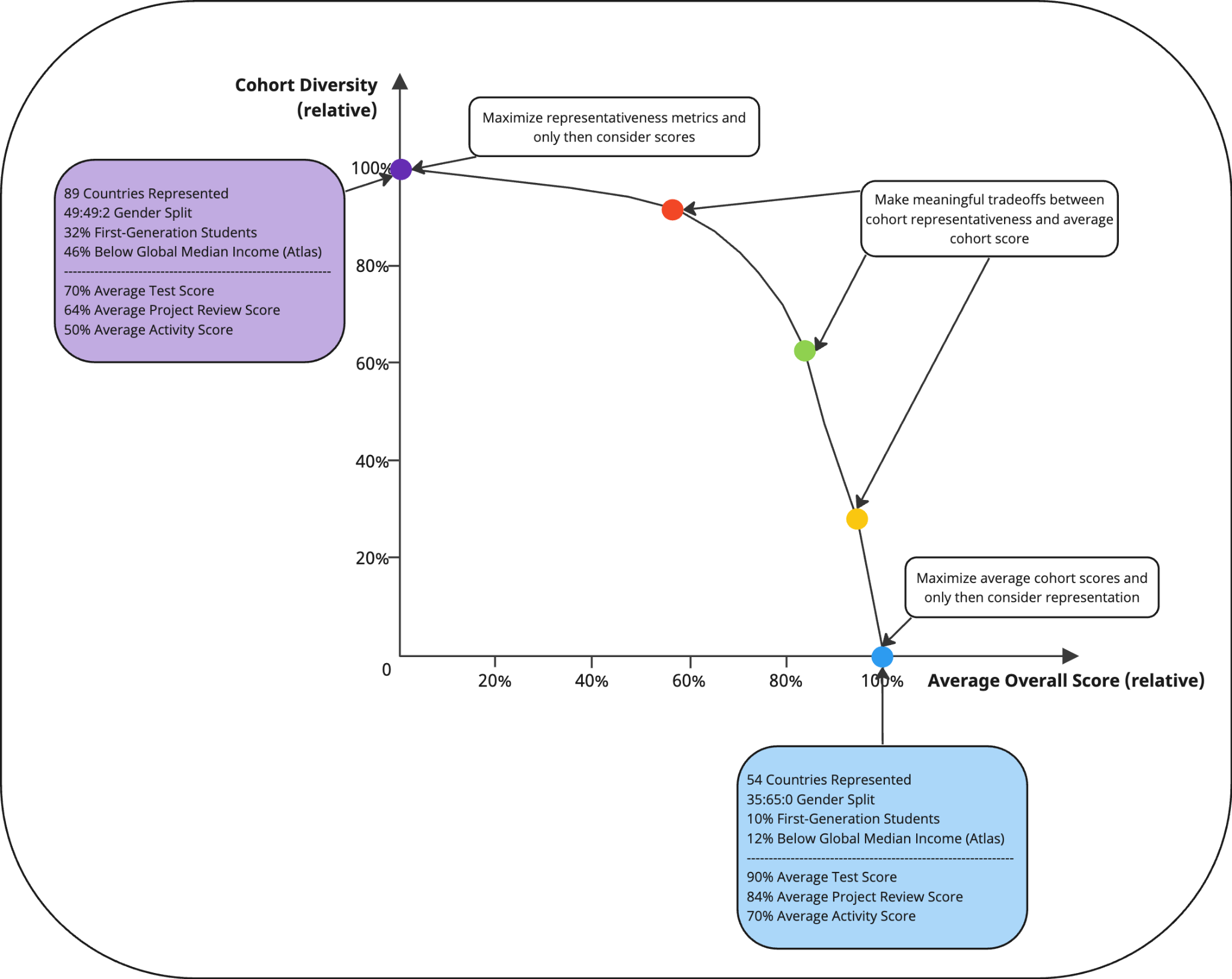}
    \caption{This figure reproduces Prototype \ref{fig:diversity} at a larger scale.}
    \label{fig:diversity_full}
\end{figure}

\begin{figure}[htbp]
    \centering
    \includegraphics[width=0.9\linewidth]{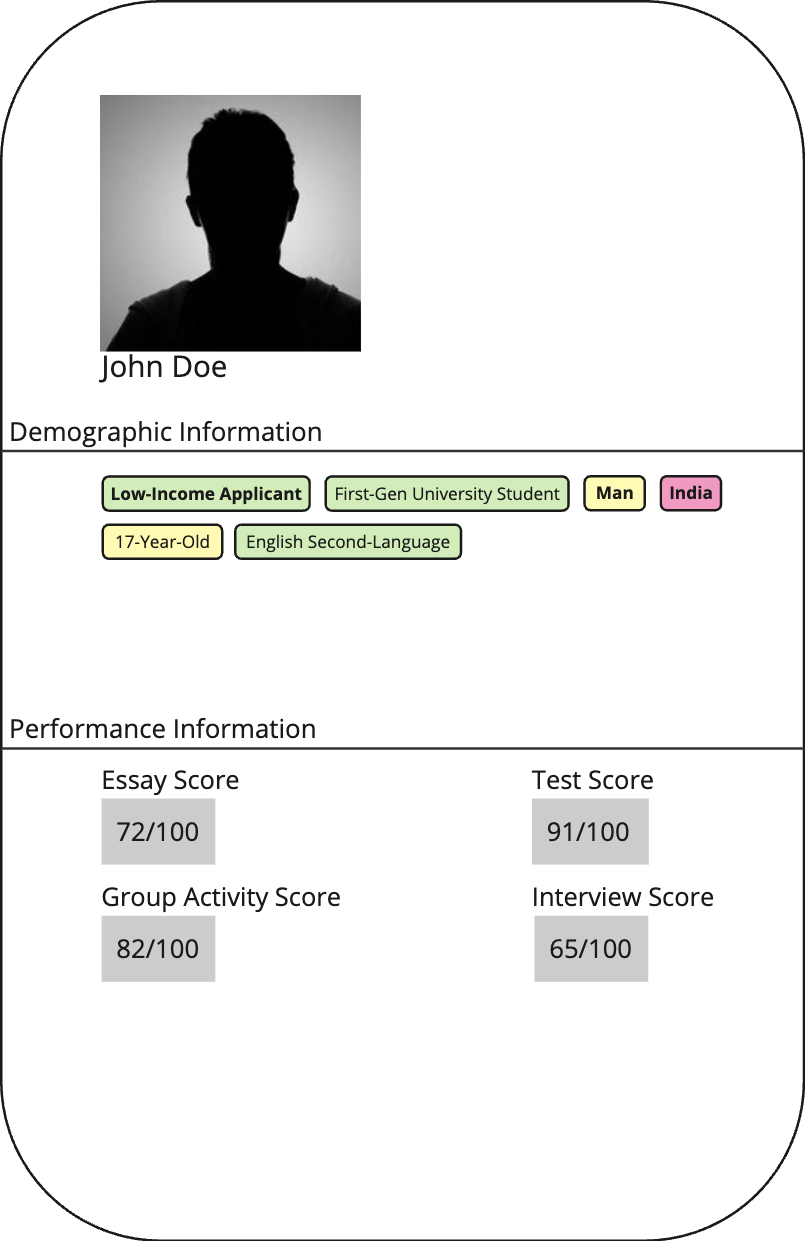}
    \caption{This figure reproduces Prototype \ref{fig:demographic} at a larger scale.}
    \label{fig:demographic_full}
\end{figure}

\begin{figure}[htbp]
    \centering
    \includegraphics[width=0.9\linewidth]{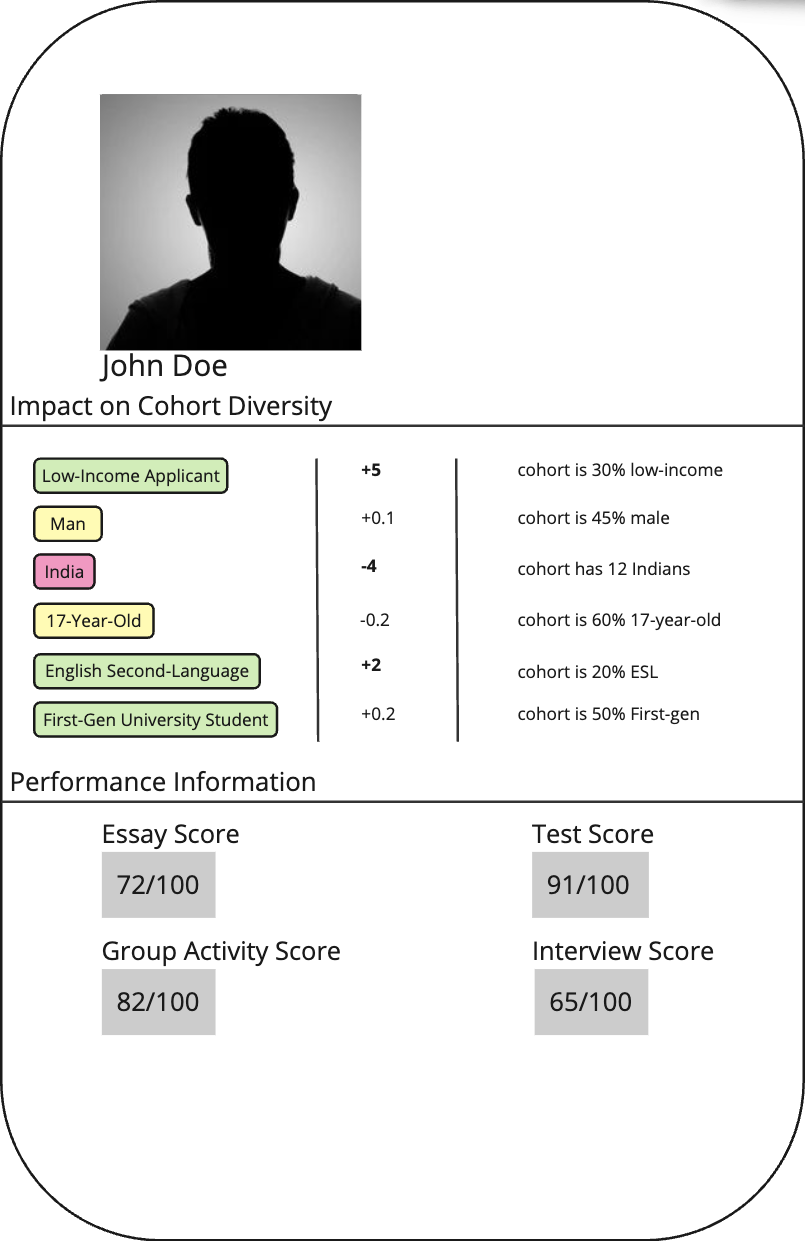}
    \caption{This figure reproduces Prototype \ref{fig:impact} at a larger scale.}
    \label{fig:impact_full}
\end{figure}

\begin{figure}[hbtp]
    \centering
    \includegraphics[width=0.9\linewidth]{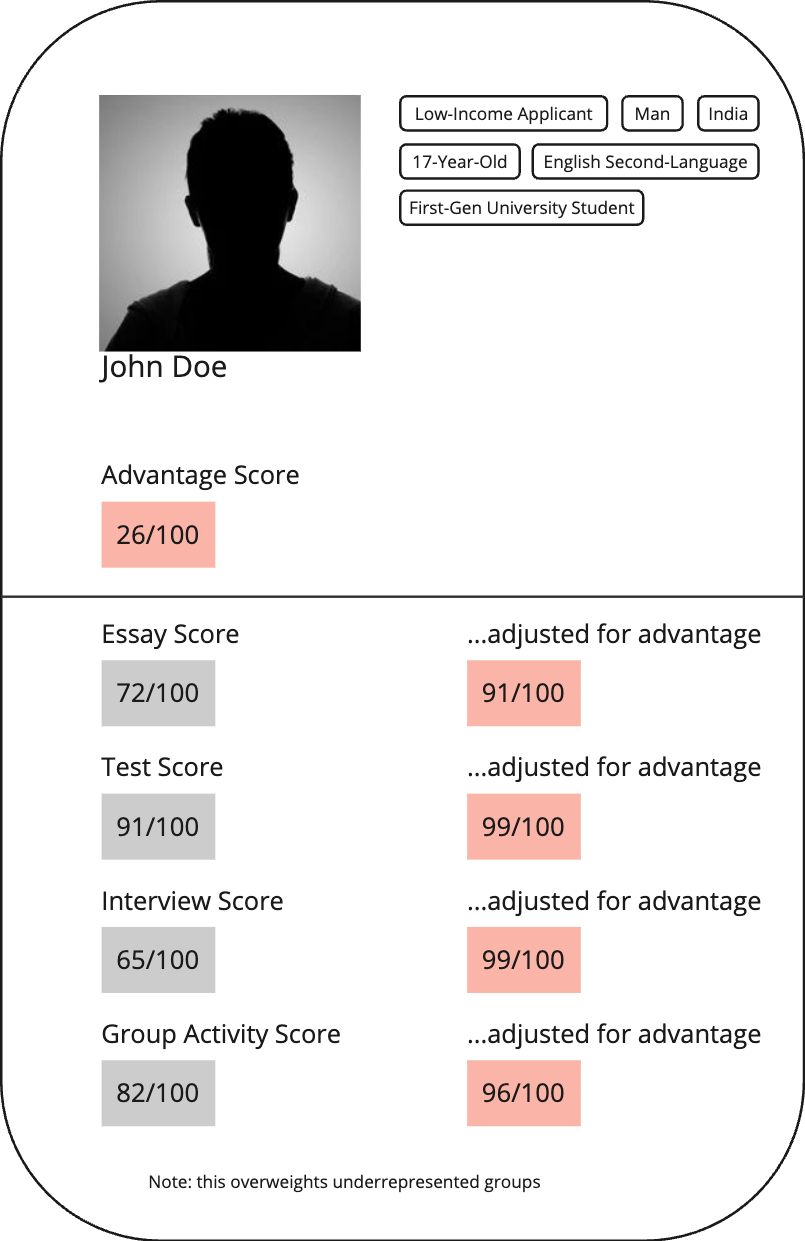}
    \caption{This figure reproduces Prototype \ref{fig:advantage} at a larger scale.}
    \label{fig:advantage_full}
\end{figure}

\end{document}